\documentclass[sigconf]{acmart}

\AtBeginMaketitle{%

\def\EDBTISSN{2367-2005}
\def\EDBTISBN{978-3-98318-104-9}
\setcopyright{none}
\copyrightyear{\copyright\ 2026 Copyright held by the owner/author(s).
  Published on \url{OpenProceedings.org} under ISBN \EDBTISBN, series ISSN \EDBTISSN.
  Distribution of this paper is permitted under the terms of the
  Creative Commons license CC-by-nc-nd 4.0}
\acmYear{2026}
\acmDOI{}
\acmISBN{}

\acmConference[EDBT '26]{Extending Database Technology}{24-27 March 2026}{Tampere (Finland)}

\settopmatter{printacmref=false, printccs=false, printfolios=false}

\newsavebox{\ximagebox}
\newlength{\ximageheight}
\newsavebox{\xglyphbox}
\newlength{\xglyphheight}
\newcommand{\xbox}[1]%
  {\savebox{\ximagebox}{#1}%
  \settoheight{\ximageheight}{\usebox{\ximagebox}}%
  \savebox{\xglyphbox}{\color{white}\char32}%
  \settoheight{\xglyphheight}{\usebox{\xglyphbox}}%
  \raisebox{\ximageheight}[0pt][0pt]{\raisebox{-\xglyphheight}[0pt][0pt]{%
    \makebox[0pt][l]{\usebox{\xglyphbox}}}}%
    \usebox{\ximagebox}%
    \raisebox{0pt}[0pt][0pt]{\makebox[0pt][r]{\usebox{\xglyphbox}}}}

\newsavebox{\LogoBox}
\sbox{\LogoBox}{\includegraphics[height=1cm]{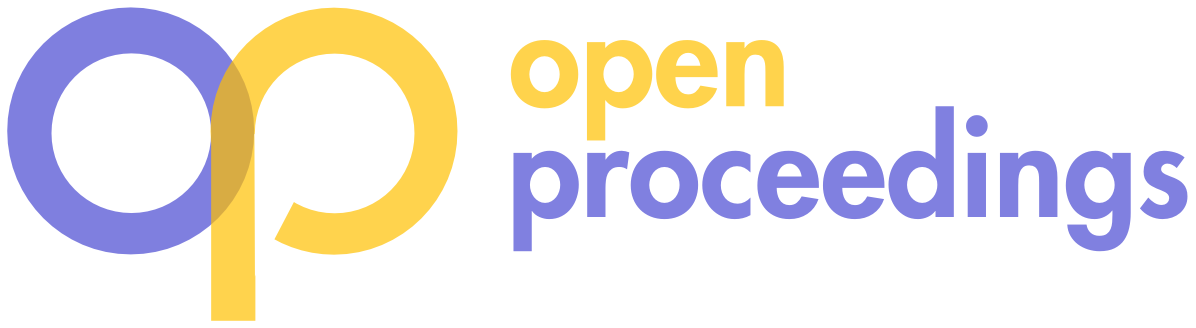}}

\fancypagestyle{firstpagestyle}{%
  \fancyhf{}%
  \fancyfoot{}%
  \fancyhead[L,C]{}
  \fancyhead[R]{\href{https://OpenProceedings.org/}{\raisebox{15pt}[0pt][0pt]{\xbox{\usebox{\LogoBox}}}}}
  }

\hypersetup{%
  pdfcopyright={CC-by-nc-nd},
  pdfpublisher={OpenProceedings.org -- University of Konstanz, Germany},
  pdfcreator={LaTeX with acmart, hyperref, and EDBT modifications},
  pdfvolumenum={29},
  pdfissuenum={2},
  pdfisbn={\EDBTISBN},
  pdfissn={\EDBTISSN}
}

}%

\newif\ifremovetext

\def\removetext#1{%
    \ifremovetext\relax\else #1\fi}
\removetexttrue

\usepackage{booktabs} %
\usepackage{graphicx}
\usepackage{comment}
\usepackage{xcolor}
\usepackage{tikz}
\usepackage{tabularx}
\usepackage{placeins}
\usepackage{pdfpages}
\usepackage{listings}
\usepackage{amsthm}
\usepackage{pdfpages}
\usepackage[ruled,vlined,linesnumbered]{algorithm2e}
\usepackage{microtype}
\usepackage{longtable}
\usepackage{pifont}
\usepackage{caption}
\usepackage{subcaption}

\begin{document}

\title{Nova: Scalable Streaming Join Placement and Parallelization in Resource-Constrained Geo-Distributed Environments}

\author{Xenofon Chatziliadis}
\email{x.chatziliadis@tu-berlin.de}
\orcid{0009-0003-1384-6603}
\affiliation{%
  \institution{BIFOLD, TU Berlin}
  \country{Germany}
}

\author{Eleni Tzirita Zacharatou}
\email{eleni.tziritazacharatou@hpi.de}
\orcid{0000-0001-8873-5455}
\affiliation{%
  \institution{HPI, Uni Potsdam}
  \country{Germany}
}

\author{Samira Akili}
\email{samira.akili@tu-berlin.de}
\orcid{0000-0002-8441-7713}
\affiliation{%
  \institution{BIFOLD, TU Berlin}
  \country{Germany}
}

\author{Alphan Eracar}
\email{eracar@campus.tu-berlin.de}
\affiliation{%
  \institution{BIFOLD, TU Berlin}
  \country{Germany}
}

\author{Volker Markl}
\email{volker.markl@tu-berlin.de}
\orcid{0009-0009-0964-026X}
\affiliation{%
  \institution{BIFOLD, TU Berlin, DFKI}
  \country{Germany}
}

\renewcommand{\shorttitle}{Nova: Scalable Streaming Join Placement and Parallelization}
\renewcommand{\shortauthors}{Chatziliadis et al.}

\begin{abstract}

Real-time data processing in large geo-distributed applications, like the Internet of Things (IoT), increasingly shifts computation from the cloud to the network edge to reduce latency and mitigate network congestion. 
In this setting, minimizing latency while avoiding node overload requires jointly optimizing operator replication and placement of operator instances, a challenge known as the Operator Placement and Replication (OPR) problem. 
OPR is NP-hard and particularly difficult to solve in large-scale, heterogeneous, and dynamic geo-distributed networks, where solutions must be scalable, resource-aware, and adaptive to changes like node failures.
Existing work on OPR has primarily focused on single-stream operators, such as filters and aggregations. However, many latency-sensitive applications, like environmental monitoring and anomaly detection, require efficient regional stream joins near data sources.

This paper introduces Nova, an optimization approach designed to address OPR for join operators that are computable on resource-constrained edge devices.
Nova relaxes the NP-hard OPR into a convex optimization problem by embedding cost metrics into a Euclidean space and partitioning joins into smaller sub-joins.
This new formulation enables linear scalability and efficient adaptation to topological changes through partial re-optimizations. We evaluate Nova through simulations on real-world topologies and on a local testbed, demonstrating up to $39\times$ latency reduction and 4.5$\times$ increase in throughput compared to existing edge-centered solutions, while also preventing node overload and maintaining near-constant re-optimization times regardless of topology size.

\removetext{Stream joins are fundamental operations for real-time data processing in geo-distributed applications, such as environmental monitoring and predictive maintenance. 
Offloading join computations to fog and edge devices can significantly reduce latency and network overhead.
However, it also presents challenges due to the constraints, heterogeneity, and volatility of these devices. 
Efficiently parallelizing join operators, along with optimizing their placement, is crucial to preventing overload in resource-constrained environments. 
Although significant research has focused on parallelizing joins in cloud environments using different strategies such as data partitioning, operator replication, and placement, there remains a significant gap in integrated solutions that address the unique challenges of geo-distributed infrastructures.
}

\removetext{
A common trend in geo-distributed stream processing is to push computation closer to data sources, as seen with filters and aggregations to reduce latency and minimize network traffic. However, applications like predictive maintenance and environmental monitoring often require joined streams before filtering, making it essential to offload join computation closer to the data sources. Identifying the placement of these joins introduces challenges due to device constraints, heterogeneity, and dynamic network conditions, which, if mismanaged, can overload nodes and increase latency. While prior research has explored placing joins in cloud environments, solutions tailored to the unique challenges of geo-distributed infrastructures remain limited.
}

\end{abstract}


\maketitle

\section{Introduction}
\label{sec:introduction}

Internet of Things (IoT) applications employ sensors in distributed locations, generating large amounts of streaming data~\cite{lepping23nesdemo, data-fusion-smart-city}.
Combining streams from different sources typically requires transmitting data to a centralized cloud~\cite{carbone2015flink, zaharia2016spark, chintapalli2016benchmarking}, introducing significant communication overhead and latency.
This is particularly inefficient when lightweight join operations based on simple predicates could instead be executed on resource-constrained edge nodes~\cite{shi2016edge, verwiebe2022algorithms, gkonis2023survey}. 
Such joins are common in real-time IoT applications, including the energy sector~\cite{rezaei2021analytical}, smart manufacturing~\cite{sahal2020big}, healthcare~\cite{healthcare-motivation}, and smart cities~\cite{smart-city-motivation, lepping23nesdemo, data-fusion-smart-city} (e.g., joining traffic and weather streams in a smart city to dynamically adjust speed limits~\cite{ziehn2025unraveling}).
In this paper, we focus on a geo-distributed environmental monitoring scenario inspired by the DEBS 2021 Grand Challenge~\cite{tahir2021debs}. The challenge studies large-scale air-quality monitoring using geographically distributed sensors, which requires correlating time-windowed measurements across regions to detect environmental changes. 
We consider a similar geo-distributed environmental sensing workload using data from Sensor.Community~\cite{sensorcommunity} where pressure and humidity streams from sensors in multiple regions are continuously joined by region identifier and time window to detect regional climate anomalies.
Executing these joins centrally quickly leads to high latency and overload, whereas placing and parallelizing them across edge and fog nodes keeps latency low while respecting resource constraints. We use this scenario as our main end-to-end use case throughout the paper and revisit it in our evaluation.

\textbf{Challenges in the Edge-Cloud Continuum.} To address the limitations of cloud-centered computation, SPEs such as NebulaStream~\cite{Zeuch2020nes} have emerged, enabling low-latency processing by distributing tasks across the entire edge-cloud continuum. This design aligns with the osmotic computing paradigm~\cite{villari2016osmotic}, which dynamically balances computation between cloud and edge resources based on network conditions and resource availability.
A key mechanism in this approach is the replication and strategic placement of computational operators (e.g., filters, joins, and aggregations) across nodes to meet objectives such as latency minimization and bottleneck avoidance.
This joint decision-making task is referred to as Operator Placement and Replication (OPR)~\cite{cardellini2018optimalrep, chatziliadis2024nemo}.
OPR is an NP-hard problem, as it generalizes the well-known General Assignment Problem~\cite{cattrysse1992surveyga}. Solving OPR requires accounting for a variety of complex factors, including network topology, resource constraints, and operator dependencies. These challenges are exacerbated in osmotic computing environments, which exhibit three key characteristics: (1) large scale, potentially encompassing millions of nodes~\cite{Zeuch2020nes}; (2) high heterogeneity, with devices ranging from resource-constrained edge nodes (e.g., Raspberry Pis) to powerful cloud servers~\cite{gao2017fog}; and (3) extreme dynamics of fluctuating data rates, and frequent changes in the network due to mobility or IP-level routing~\cite{da2022fog}. As a result, effective OPR strategies must be scalable, resource-aware, robust to frequent changes, and capable of supporting partial re-optimization~\cite{gkonis2023survey, dauda2024survey}.

\textbf{Limitations of Existing Work.}
Existing work on OPR in osmotic computing mainly spans two domains: distributed stream processing (DSP) and wireless sensor networks (WSNs). In DSP, OPR is typically solved using Integer Linear Programming (ILP), which scales poorly and requires full re-computation after system changes~\cite{cardellini2018optimalrep}.
In contrast, WSN research focuses on adaptive methods that parallelize tuple-wise joins across clustered or tree-based overlays toward the nearest base station~\cite{mamun2012wsnsurvey, WSN_SURVEY}. However, these approaches are unsuitable for DSP workloads, as they cannot span across geo-distributed deployments and are resource-agnostic, leading to node overloading and high latencies~\cite{mamun2012wsnsurvey, WSN_SURVEY}.

Recently, we introduced NEMO~\cite{chatziliadis2024nemo}, a scalable, resource-aware, and adaptive solution for the placement of decomposable aggregation operators in osmotic environments.
Extending NEMO to joins is non-trivial for three reasons:
(1) Joins amplify data rather than reduce it. NEMO parallelizes aggregation by computing partial results hierarchically, reducing data volume at each level. Joins, however, can produce outputs larger than their inputs—in the worst case, a cross-product—making this tree-based parallelization inapplicable.
(2) Parallelizing aggregation requires only partitioning the input stream. Parallelizing joins, however, requires partitioning one input while broadcasting the other to all replicas, creating a trade-off between computational load and network traffic that NEMO does not address.
(3) Unlike aggregations, joins cannot be computed hierarchically. Each replica operates independently, requiring a different placement approach.
As a result, a robust, resource-aware, adaptive, and scalable solution to the OPR problem for join operators in DSP remains an open challenge.

~\textbf{Contributions.}
This paper presents Nova, an optimization approach for streaming joins in heterogeneous, geo-distributed environments.
Building on the cost-space abstraction of NEMO, Nova introduces: (1) a cost model to handle data amplification, (2) a bandwidth-aware partitioning strategy to control network overhead, and (3) a geometric median based placement approach to optimize per-partition latency for independent replicas. 
Nova jointly optimizes placement, replication degree, and stream partitioning for 2-way joins, achieving linear-time scalability, rapid re-optimization, and resilience to dynamic network behavior. Specific contributions are as follows:

\begin{enumerate}
    \item \textit{Problem Definition}: 
    Section~\ref{sec:preliminaries} introduces our stream processing model and formalizes the Operator Placement and Parallelization (OPP) problem. OPP extends OPR with the identification of optimal stream partitioning, which is required to solve OPR for join operators.

    \item \textit{Nova Optimization Framework}: Section~\ref{sec:nova} presents our core contributions: (1) To address data amplification, we introduce a decomposition strategy based on the join matrix that partitions joins into sub-joins according to data locality and resource availability, enabling horizontal parallelization near sources. (2) To handle the independence of join replicas, we formulate placement as a convex geometric median problem. (3) To satisfy bandwidth constraints, we develop a partitioning scheme that jointly optimizes replication and network traffic. (4) To support dynamic environments, we present a re-optimization strategy that adapts to topology changes and workload shifts without full recomputation of the placement. Collectively, these techniques address optimization challenges inherent to join placement and parallelization in resource-constrained geo-distributed environments.

    \item \textit{Experiments}: In Section~\ref{sec:evaluation}, we compare Nova against heuristics used in SPEs and adaptive join techniques used in WSNs. Through simulations, we demonstrate that Nova remains efficient even on topologies with up to a million nodes, while supporting near-constant re-optimizations in under one second. These results highlight Nova's potential as an optimization approach for future SPEs designed to operate at this scale.
    Furthermore, we integrate Nova into NebulaStream's optimizer~\cite{Zeuch2020nes} and evaluate Nova's placement performance on a local heterogeneous computing cluster.
    Unlike state-of-the-art approaches that are resource-oblivious and may overload nodes, Nova avoids over-utilization entirely. Nova achieves hereby latency improvements of 4.2–39$\times$ and 4.5$\times$ higher throughput compared to existing approaches.
\end{enumerate}

\section{Preliminaries}

\label{sec:preliminaries}
This section introduces the fundamental concepts and definitions of the Nova approach.
We first define the semantics of a stream processing application and stream join in Section~\ref{sec:preliminaries_spe_model}.
Next, we describe how Nova models latency and resource metrics for placement decisions in Section~\ref{sec:preliminaries_resource_model}.
Finally, in Section~\ref{sec:preliminaries_problem}, we formulate the operator placement and parallelization problem for join operators that Nova addresses.

\subsection{Stream Processing Model}
\label{sec:preliminaries_spe_model}
A stream processing application typically consists of one or more user-defined queries compiled into a directed graph of operators interconnected by data streams.
Operators are self-contained units performing specific functions, while streams are unbounded sequences of data tuples. To enable meaningful computations, data streams are typically discretized using window operators~\cite{traub2019}.

\textbf{Stream Model.} Let $\Omega$ be the universe of operators. 
Each operator $\omega = (\text{id}, r, \rho, L_{in}, L_{out}) \in \Omega$ is characterized by several attributes. 
The attribute $\text{id}$ represents the operator's unique identifier, while $r$ denotes its replication number. 
Together, $\text{id}$ and $r$ form a unique identifier for each operator instance. 
The attribute $\rho$ specifies the total number of operator instances that share the same $\text{id}$.
The set of incoming streams is defined by $L_{in} = \{s_1, \dots, s_n\}$, whereas the set of outgoing streams is given by $L_{out} = \{t_1, \dots, t_n\}$. 
Operators that exclusively produce streams, i.e., those for which $L_{in} = \emptyset$, are called \emph{sources}, whereas those that only consume streams, i.e., those for which $L_{out} = \emptyset$, are referred to as \emph{sinks}.
Two operators in a given $\Omega$ are considered connected if the output stream of one operator serves as the input stream of another. 
We denote the set of connected operators in $\Omega$ as  $\mathit{con}(\Omega)$
$ = \{(\omega_i, \omega_j) \in \Omega^2 \mid \exists t \in \omega_i.L_{out} \ [t \in \omega_j.L_{in}]\}.$  
The dot notation provides attribute access, e.g, $\omega.L_{in}$, represents the incoming streams of the operator $\omega$.

\textbf{Logical Plan.} We formally define the initial input plan representing a DSP application as a logical plan $\Omega_{log} \subseteq \Omega$. 
Each operator $\omega_{log} \in \Omega_{log}$ has a replication factor of one, i.e., $\omega_{log}.\rho = 1$.

\textbf{Stream Join.} The stream join operator combines tuples from two streams based on a join condition. Given two streams $i$ and $j$, a stream join operator produces a new stream $ {i \Join j} $ with tuples that satisfy a join condition $ \Theta $. Formally, the join is expressed as: 

\[
i \Join j = \{ (x, y) \mid x \in i, y \in j, \Theta(x, y) \}.
\]

\begin{figure}[t!]
  \includegraphics[width=\columnwidth]{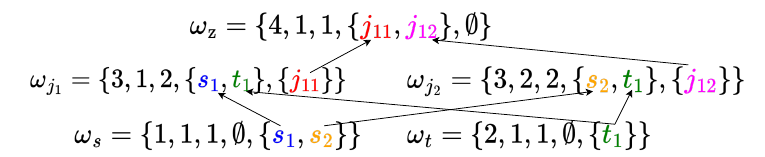}
  \caption{Formalization of a decomposable join $s \Join t$, where each join replica $\omega_{j_1}, \omega_{j_2}$ processes a partition of streams $s$ and $t$, forwarding partial results to the sink.}
  \label{fig:formalization_join}
\end{figure}

\noindent The join condition $ \Theta $ in our work represents an operation with low computational complexity, e.g., an equality join with $ x.A = y.B $. We note that a stream join over windows requires tuples to satisfy both the join condition and window boundary constraints, such as overlapping timestamps for time-based windows.
We illustrate our model for joining streams $s \Join t$ in~\autoref{fig:formalization_join}.

\textbf{Join Matrix.} We represent permissible joins between stream partitions \( S = \{ s_1, \dots, s_m \} \) and \( T = \{ t_1, \dots, t_n \} \) using a binary \( m \times n \) matrix \( M \), where \( M_{p,q} = 1 \) means \( s_p \) can join with \( t_q \). For predefined conditions (e.g., joins on spatial region identifiers), \( M \) is known beforehand. If join validity is uncertain, \( M \) is initialized as dense (all ones), requiring evaluation of all stream pairs. Section~\ref{sec:nova_extensions} explains how Nova handles uncertainties and updates to \( M \) at runtime. 
Importantly, \( M \) indicates only which stream partitions can join; tuple-level validity must still be checked during execution. Since we focus on two-way joins, \( M \) captures only direct joinability without considering transitive joins or join ordering.

\subsection{Resource Model}
\label{sec:preliminaries_resource_model}
We model the topology as a directed graph $G_T = (V, E)$, where $V$ is the set of nodes (e.g., servers or sensors) and $E$ the communication links between them.
The processing cost of a streaming join operator depends primarily on input data rates, window size, and join predicate complexity, than on the window type~\cite{gulisano2017performance, ziehn2025unraveling}.
On resource-constrained edge devices, this behavior simplifies further. Our empirical results for edge-computable lightweight joins (e.g., equi-joins) indicate that compute demand is mainly driven by tuple arrival rate, whereas end-to-end delay is dominated by network latency (cf. Section~\ref{sec:eval_nova_deployment}).
We therefore adopt the following model.
For an operator $\omega$ with input streams $L_{in}$ and per-stream data rates $dr(\cdot)$, the required compute capacity is
$
C_r(\omega)=\sum_{s \in L_{in}(\omega)} dr(s).
$
This capacity is benchmarked per node type and operator class in advance. Window size and predicate complexity remain fixed for a given $\Omega_{log}$ and are omitted from the notation for brevity.

Path delay is approximated by the sum of link latencies along the route, as millisecond-scale network delays dominate the sub-microsecond processing time of lightweight joins.
A placement is feasible only if $C_r(\omega) \le C_a(\nu)$ for every replica on node $\nu$ and if the aggregate traffic on each link does not exceed $b(\nu_i,\nu_j)$ when bandwidth budgets are enforced
(cf. Eqs.~\ref{eq:const_capcon} and \ref{eq:const_bw}).
Nova is agnostic to whether links are physical paths or logical channels because the cost-space embedding (cf. Section~\ref{sec:nova_embedding}) abstracts routing details.

\subsection{Problem Formulation}
\label{sec:preliminaries_problem}
\textbf{Operator Placement.} Operator placement (OP) maps each operator $\omega \in \Omega_{log}$ to a node $\nu \in V$ in a topology using a mapping function $f_m \colon  \Omega \mapsto V$~\cite{cardellini2016op}. 
Each operator must be assigned to exactly one node, which is enforced by the exclusivity constraint, i.e., $\forall \omega \in \Omega_{log} \ [\exists! \nu_j \in V \ [f_m(\omega) = \nu_j ]]$. 
Some operators, like sources and sinks, are fixed to specific nodes due to dependencies, while others, like joins, can be placed flexibly across nodes.
In the classical OP formulation, the optimization objective can vary (e.g., minimizing bandwidth usage, balancing load). In this work, we focus on latency minimization as the primary optimization objective.

\textbf{Operator Placement and Replication.} \citet{cardellini2018optimalrep} show that the general operator placement problem can be extended to the operator placement and replication problem (OPR), which jointly determines both the placement of operators and their degree of replication to optimize a given objective while satisfying resource constraints. Since OPR generalizes OP, it remains NP-hard.

\textbf{Operator Placement and Parallelization.} In this work, we further extend OPR by considering \emph{stream partitioning} alongside placement and replication. We define this as the Operator Placement and Parallelization problem (OPP). While OPR optimizes placement and replication of whole operators, OPP enables fine-grained parallelization at the stream level via partitioning, allowing a single operator’s input stream to be divided across multiple replicas.

In OPP, parallelization is controlled by a function $f_p \colon \Omega_{log} \mapsto 2^{\Omega}$, which maps each operator $\omega_{log} \in \Omega_{log}$ to a set of parallel instances ${\omega_1', \dots, \omega_\rho'}$. Each instance processes a partition of the operator’s input/output streams, ensuring $\bigcup_{\omega' \in f_p(\omega_{log})} \omega'.L_{out} = \omega_{log}.L_{out}$.
Each instance of $f_p$ defines a parallelization strategy, yielding a parallelized logical plan $\Omega'_{log} = \bigcup{\omega_{log} \in \Omega_{log}} f_p(\omega_{log})$. Parallelization reduces resource demands per replica, since partitions lower input data rates ($dr(l') \leq dr(l)$).
As OPP generalizes OPR through additional partitioning, the problem remains NP-hard.

\textbf{Optimization Objective.} Given a logical plan $\Omega_{log}$ and a topology $G_T = (V, E)$, the goal of OPP is to jointly determine: 
\begin{enumerate}
    \item A parallelization strategy $f_p$, which defines a parallelized logical plan $\Omega'_{log} \in 2^{\Omega}$, selecting a subset of operator replicas from the full search space $\Omega$.
    \item A placement strategy $f_m$, mapping each operator replica $\omega' \in \Omega'_{log}$ to a node $\nu \in V$.
\end{enumerate}
The objective is to select $f_p$ and $f_m$ such that 1) all constraints are satisfied, and 2) the total latency between connected operators is minimized.
Formally, OPP can be expressed as finding:

\begin{equation}
\label{eq:opt_func}
\Omega_{log}^\star = \underset{\Omega'_{log} \subseteq 2^{\Omega}}{\arg\min} \; \sum_{(\omega_i', \omega_j') \in con(\Omega'_{log})} d(f_m(\omega_i'), f_m(\omega_j')),
\end{equation}

\noindent subject to the following constraints:

\paragraph{1) Resource Capacity Constraint} 
The required capacity of operators must not exceed their nodes' available resources:

\begin{equation}
\label{eq:const_capcon}
\begin{aligned}
    C_r(\omega') \leq C_a(f_m(\omega')), \quad \forall \omega' \in \Omega_{log}^\star.
\end{aligned}
\end{equation}

\paragraph{2) Resource Availability Constraint} 
The available resources of assignable nodes must be above a threshold $C_{min}$.

\begin{equation}
\label{eq:const_capmin}
\begin{aligned}
    C_a(\nu_i) \ge C_{min}, \quad \forall \nu_i \in \mathit{img}(f_m).
\end{aligned}
\end{equation}

\paragraph{3) Network Bandwidth Constraint} 
Since the required capacity $C_r(\omega')$ is defined as the sum of incoming data rates, it also represents the bandwidth utilization.
Thus, the bandwidth demand must remain within the available bandwidth threshold $t_b$:

\begin{equation}
\label{eq:const_bw}
C_r(\omega') \leq t_b, \quad \forall \omega' \in \Omega_{log}^\star.
\end{equation}

This OPP formulation minimizes the sum of all end-to-end latencies while respecting constraints. An alternative objective would be to minimize the maximum end-to-end latency across partitions, corresponding to a min--max relay placement problem that optimizes only the slowest partition. We choose the min--sum objective for three reasons. First, it yields a geometric median formulation with a unique and stable optimum that can be computed efficiently. Second, it optimizes aggregate latency across all partitions rather than focusing solely on a single worst-case path. Third, it is significantly more robust to noise, estimation errors, and latency fluctuations in dynamic networks, since a single inaccurate or stale measurement can dominate a min--max objective.

\section{The Nova Approach}
\begin{figure}[t!]
  \includegraphics[width=\columnwidth]{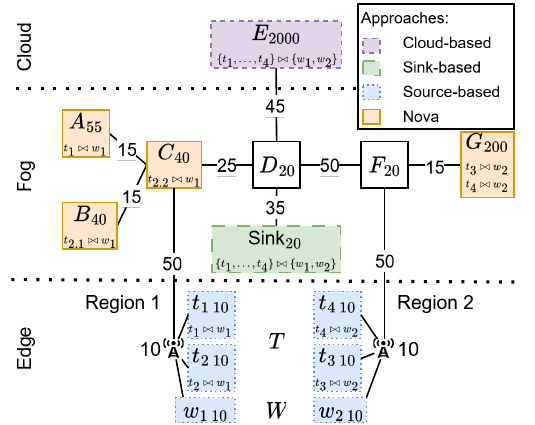}
  \caption{Running example to demonstrate Nova for joining geo-distributed sensor readings of $\{t_1, \dots, t_4\} \in T$ with  $\{w_1, w_2\} \in W$. Subscripts indicate node processing capacity, and edge labels denote network path latency.}
  \label{fig:join_strategies}
\end{figure}

\label{sec:nova}
This section introduces Nova, a scalable, adaptive, and resource-aware solution to the OPP problem for join operators in resource-constrained distributed stream processing.
Algorithm~\ref{alg:nova} outlines Nova’s workflow: Given the topology $G_T$, a logical plan $\Omega_{log}$, and the join matrix $M$, Nova produces an operator-to-node mapping for a parallelized logical plan $\Omega'_{log}$, including replicated operators with partitioned streams.
Since solving the OPP over a \textit{discrete} set of nodes is NP-hard (cf. Section~\ref{sec:preliminaries_problem}), Nova instead reformulates the problem in a continuous space and tackles it by decomposing it into three heuristic phases, each solvable in linear time.

\begin{enumerate}
    \item \textbf{Phase I: Cost Space Construction}: The discrete topology is embedded into a continuous cost space by assigning each node $\nu \in V$ a coordinate vector $\hat{\nu} \in \mathbb{R}^d$ (line 2). This mapping enables an iterative approximation of the OPP in the continuous domain, which is required for Phase II. Section~\ref{sec:nova_embedding} provides details on Phase I.
    
    \item \textbf{Phase II: Virtual Join Placement}: In the cost space, the optimization objective of OPP reduces to identifying the geometric median, a convex problem that can be efficiently solved using iterative algorithms. This formulation avoids combinatorial search over discrete placements and enables efficient approximation of optimal operator positions (lines 3--4). Section~\ref{sec:nova_vp} describes Phase II.
    
    \item \textbf{Phase III: Physical Replica Assignment}: After placement in continuous space, Nova maps the virtual solution back to the discrete topology while enforcing all resource constraints (lines 5--13). Section~\ref{sec:nova_placement} describes Phase III.
\end{enumerate}

To handle dynamic changes (e.g., node failures), Nova leverages the convexity of the cost space objective, enabling local re-optimization of affected operators without full recomputation. We discuss the re-optimization of Nova in Section~\ref{sec:nova_reopt}. While Nova’s cost space models latency by default, it is metric-agnostic and adaptable to other objectives. In Section~\ref{sec:nova_extensions} we discuss possible extensions.

\subsection{Running Example}
\label{sec:nova_example}
For clarity, we use a simplified environmental monitoring workload, shown in Fig.~\ref{fig:join_strategies}, as a running example throughout the rest of Section~\ref{sec:nova}. Two sensor streams, a pressure stream $T=\{t_1,t_2,t_3,t_4\}$ and a humidity stream $W=\{w_1,w_2\}$, are generated by geographically distributed \textit{Sensor.Community} nodes~\cite{sensorcommunity} and joined on region identifier and time window before being forwarded to a local sink to detect regional climate anomalies.
The topology follows an edge–fog–cloud pattern with sources at the edge, several medium-capacity fog nodes (A–G), a high-capacity cloud node (E), and a local sink. 
Each source emits at $25\,\text{Hz}$, edge labels denote network latency (ms), and node subscripts indicate processing capacity (tuples/s). 
The join decomposes into two region-specific sub-joins:
$
T \Join W = \big( (t_1 \Join w_1) \cup (t_2 \Join w_1) \big) \cup \big( (t_3 \Join w_2) \cup (t_4 \Join w_2) \big).
$

We contrast Nova with three baselines.
In a source-based or sink-based placement, all six sources together generate $6\times 25=150$ tuples/s, but each source processes only $10$ tuples/s and the sink $20$ tuples/s, leading to overload and backpressure.
A cloud-based strategy places the join on node E, where Region~1 traffic must traverse C and D, and Region~2 traffic F and D, which yields path delays of roughly $130$\,ms and $155$\,ms, respectively. Returning results to the sink adds about $100$\,ms, so the maximal end-to-end latency reaches about $275$\,ms.
Nova decomposes the join into two region-specific sub-joins, placing Region~2's sub-join on node G and parallelizing Region~1's sub-join across nodes A, B, and C by broadcasting $w_1$ and partitioning $t_2$ into $t_{21}$ and $t_{22}$.  
This ensures that for each operator $\omega$ on node $\nu$, the condition $C_r(\omega) \le C_a(\nu)$ is met, reducing end-to-end latency to roughly $150$\,ms for Region~1 and $175$\,ms for Region~2.
The following subsections explain how Nova derives this plan.

\begin{algorithm}[t!]
\SetAlgoLined
\SetKwInOut{Input}{Input}
\SetKwComment{Comment}{// }{}

\Input{$G_T$ (topology), $\Omega_{log}$(logical plan), $M$ (join matrix)}
$placements \gets \{ \}$\;
$\hat{V} \gets \text{compute\_coordinates}(G_T)$\;

$\Omega_{log}'\gets \text{resolve\_operators}(\Omega_{log}, M)$\;
$\hat{\Omega} \gets \text{compute\_optima}(\hat{V}, \Omega_{log}')$\;

\ForEach{$\omega' \in \Omega_{log}'$}{
    \If{$ \text{is\_pinned}(\omega')$}{
        $placements[\omega'] \gets \text{get\_pinned\_node}(\omega')$\;
    }
    \Else{
        $result \gets \text{parallelize\_and\_place}(\omega', \Omega_{log}', \hat{V}, \hat{\Omega})$\;
        $placements.\text{update}(result)$\;
    }
}
\Return $placements, \Omega_{log}'$

\caption{The Nova approach}
\label{alg:nova}

\end{algorithm}

\subsection{Cost Space Construction}
\label{sec:nova_embedding}
Following NEMO~\cite{chatziliadis2024nemo, terhaag2025gpu}, the first phase is a pre-processing stage that maps the discrete network topology to a continuous cost space (Algorithm~\ref{alg:nova}, line~2).
Conceptually, Nova represents network latencies by a symmetric matrix $A \in \mathbb{R}^{n \times n}$, where each entry $A_{ij}$ denotes the end-to-end latency between nodes $\nu_i$ and $\nu_j$.
The corresponding embedding can be formulated as the multidimensional scaling (MDS) problem of finding an embedding matrix $\hat{G}_T \in \mathbb{R}^{n \times d}$ whose induced distance matrix $D(\hat{G}_T)$ approximates $A$ by minimizing
\begin{equation}
    \min_{\hat{G}_T \in \mathbb{R}^{n \times d}} \|D(\hat{G}_T) - A \|_F^2 .
\end{equation}

Here, each row of $\hat{G}_T$ represents the coordinates of a node in the cost space, with the $i$th row of $\hat{G}_T$ corresponding to the coordinates of the $i$th node $\nu_i \in V$.
We denote the coordinates of node $\nu_i$ as $\hat{\nu}_i$ and represent the set of all node coordinates as $\hat{V} = \{\hat{\nu}_1, \dots, \hat{\nu}_n\}$.

For very large latency matrices \(A\), Nova does not solve the dense MDS formulation. Instead, Nova computes node coordinates using Vivaldi~\cite{vivaldi04}, which maintains a small neighbor set \(\mathcal{N}(i)\) for each node \(\nu_i\) and materializes latency entries only for pairs \((i,j)\) with \(j \in \mathcal{N}(i)\), where \(|\mathcal{N}(i)| = m \ll |V|\). Vivaldi thus acts as a stochastic solver for the above MDS objective over this neighborhood-induced sparse distance matrix and directly produces the coordinate set \(\hat{V} = \{\hat{\nu}_1, \dots, \hat{\nu}_n\}\) used by Nova, while avoiding quadratic measurement and computation overhead.

\textbf{Benefits.} Nova relies on the cost space to efficiently solve the OPP using heuristic methods. Beyond enabling placement optimization, the embedding offers several key advantages: 1) It abstracts physical network paths, making it well-suited for geo-distributed systems with unknown or dynamic routing. 2) It increases robustness to continuous latency fluctuations, reducing the need for frequent re-optimization (cf. Section~\ref{sec:eval_evolution}). 3) It significantly reduces the number of latency measurements and makes the cost-space construction scalable to large topologies.

\textbf{Limitations.} Euclidean embeddings assume that network latencies satisfy the triangle inequality, but real-world Internet latencies often violate this property, a phenomenon known as Triangle Inequality Violations (TIVs)~\cite{ledlie2007network}. Like other NCS-based approaches~\cite{vivaldi04, PietzuchNAOP}, Nova’s placement quality depends on the accuracy of estimated latencies, and severe TIVs can reduce placement quality. We evaluate the practical impact of TIVs in Section~\ref{sec:eval_estimation_error} by comparing estimated and measured latencies across real-world topologies. 
In practice, geo-distributed topologies are highly dynamic, and even exact measurements quickly become stale due to routing and network changes. Nova, therefore, emphasizes robustness to moderate estimation errors and continuous adaptation rather than strict accuracy.

\begin{figure*}[t!]
  \subfloat[Virtual Placement]{\includegraphics[scale=0.4]{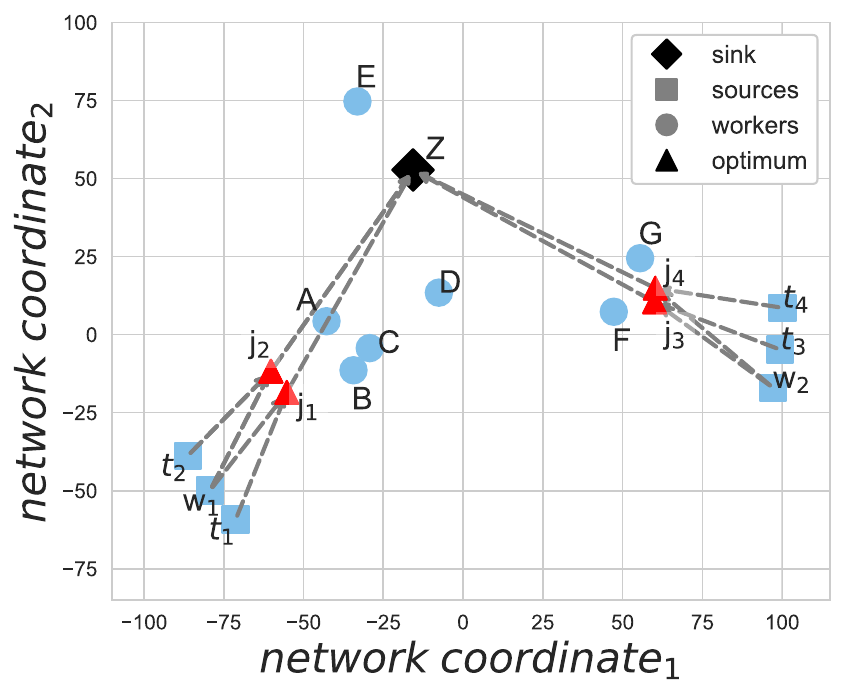}}
  \subfloat[Physical Placement]{\includegraphics[scale=0.4]{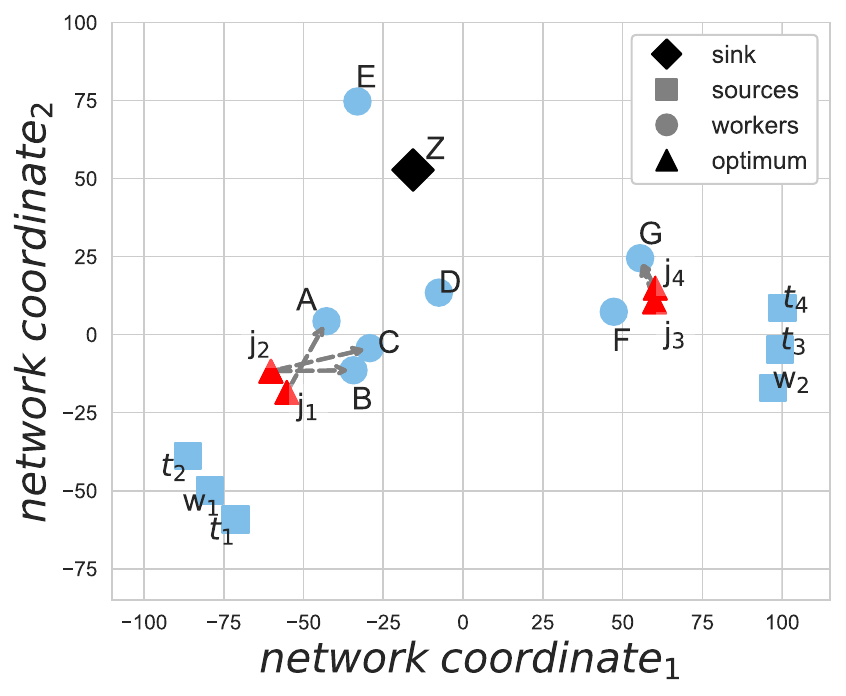}}
  \subfloat[Final Placement]{\includegraphics[scale=0.4]{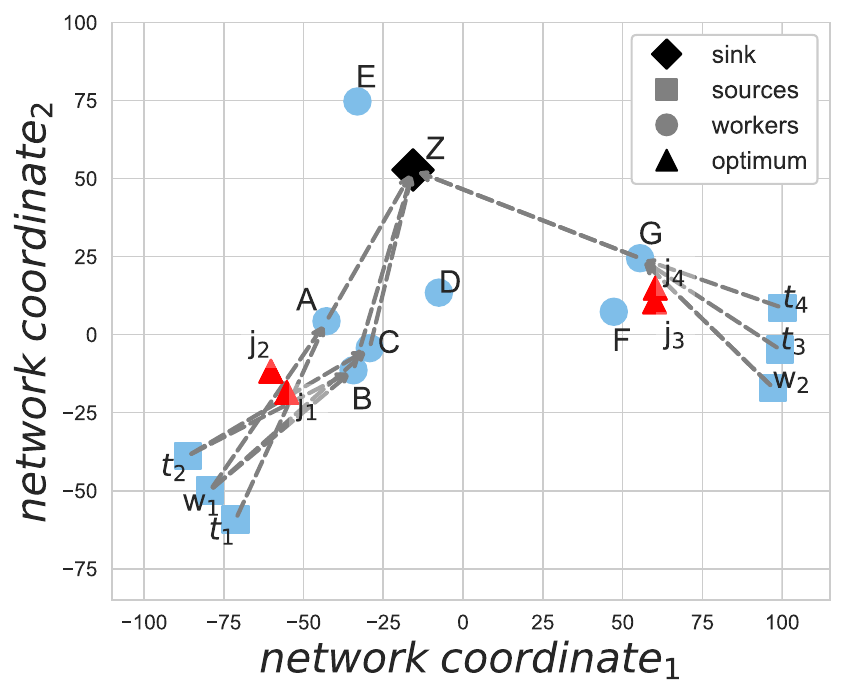}}
  \caption{Progression of Nova illustrated on the example of Section~\ref{sec:nova_example}, where the topology is embedded in a $\mathbb{R}^2$ Euclidean space.}
  \label{fig:nova_overview}
\end{figure*}

\textbf{Example.} Figure~\ref{fig:nova_overview} illustrates the Euclidean embedding of the example topology, with worker nodes shown as~$\bullet$, sources as~$\blacksquare$, and the sink as~$\blacklozenge$. The embedding is derived from the measured latencies in Figure~\ref{fig:join_strategies}. For instance, the latency from $t_1$ to $C$ is $60$\,ms ($10$\,ms to the base station and $50$\,ms to $C$), whereas the end-to-end latency from $t_1$ to the sink is $110$\,ms. Accordingly, the matrix~$A$ contains entries $A[t_1, C] = 60$ and $A[t_1, \text{sink}] = 110$. In the Euclidean space, $t_1$ and $C$ are therefore placed close together, while $t_1$ appears farther from the sink. These distances guide the continuous optimization of the join operator locations (shown as~$\blacktriangle$).

\begin{figure}[t!]
  \centering
  \subfloat[Input]{\includegraphics[width=0.19\linewidth]{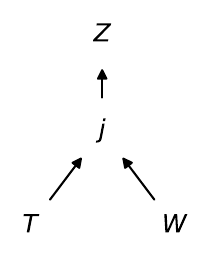}}
  \hspace{0.01\linewidth}
  \subfloat[Resolved operators]{\includegraphics[width=0.38\linewidth]{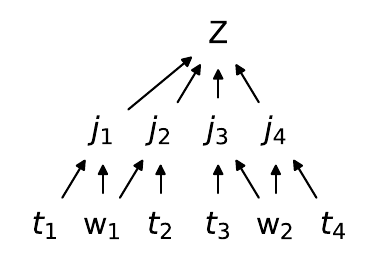}}
  \hspace{0.01\linewidth}
  \subfloat[Partitioned streams]{\includegraphics[width=0.38\linewidth]{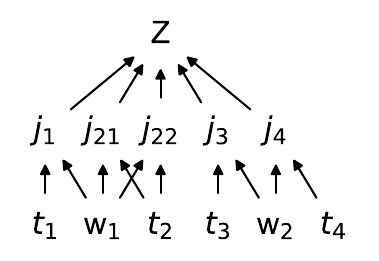}}
  \caption{Different stages of the logical operator plan in Nova. }

  \label{fig:logical_plans}
\end{figure}

\subsection{Virtual Join Placement}
\label{sec:nova_vp}
In the second phase, Nova efficiently approximates the OPP problem by first solving it in the continuous cost space.
The process consists of three steps: 1) source expansion, 2) pair-wise join replication, and 3) computation of the optimal placements of the resolved operators in the cost space. Steps 1 and 2 together form Nova's \textit{resolving operators} function (cf. Algorithm~\ref{alg:nova}, line~3).

\textbf{Source Expansion.} Nova begins by expanding the logical source streams in $\Omega_{log}$ into their corresponding physical streams located at data-producing nodes. 
A single logical stream may comprise arbitrarily many physical streams sharing the same schema. Thus, even if a join has only two logical inputs, the underlying number of physical streams can be very large.

\textbf{Pair-wise Join Replication.} Next, Nova creates replicas of the initial join operator $\omega_j \in \Omega_{log}$, one for each join pair defined by the join matrix $M$ (cf. Section~\ref{sec:preliminaries_spe_model}). Each replica corresponds to a specific pair of input partitions to be joined, resulting in an intermediate parallelized logical plan $\Omega'_{log}$. 

\textbf{Placement in the Cost Space.}  
After resolving the logical streams, Nova computes the placement of the join replicas in the cost space (Algorithm~\ref{alg:nova}, line~4).
This step conceptually resembles the virtual placement stage of NEMO~\cite{chatziliadis2024nemo}, where operator locations are determined by minimizing a spring-energy objective.
However, Nova's formulation differs significantly due to the structural properties of operator graphs for join replicas.
In NEMO, aggregation operators form a hierarchy, necessitating coupled optimization where placement decisions at one level influence those at other levels. In contrast, join replicas are independent: each connects only to its two sources and the sink, with no inter-replica dependencies. 
As a result, the energy is decoupled across the replicas, and the objective reduces to the geometric median in the cost space~\cite{fekete2005continuous}.

Formally, for each join replica $\omega_j \in \Omega'_{log}$, Nova considers the set of coordinates $\hat{X} = \{\hat{\nu_1}, \dots, \hat{\nu_n}\}$ representing the pinned nodes (i.e., sources and sinks) connected to $\omega_j$. The goal is to find the virtual position $\hat{\omega}_j \in \mathbb{R}^d$ that minimizes the sum of distances to all nodes in $\hat{X}$. This position corresponds to each join replicas' geometric median, which is defined as:

\begin{equation}
    \hat{\omega}_j = \underset{y \in \mathbb{R}^d}{\arg\min} \sum_{\hat{\nu} \in \hat{X}} d(\hat{\nu}, y)
\end{equation}

The geometric median is a convex optimization problem, which we solve iteratively using gradient descent~\cite{ruder2016gradient}. 
We denote the optimal virtual position of each operator $\omega_i$ as $\hat{\omega}_i$, and the set of all virtual placements as $\hat{\Omega} = \{\hat{\omega}_1, \dots, \hat{\omega}_n\}$.

\textbf{Example.} In the first step (source expansion), Nova replaces the logical sources $T$ and $W$ (Figure~\ref{fig:logical_plans}a) with their physical data-producing nodes, expanding $T$ into ${t_1,\dots,t_4}$ and $W$ into ${w_1,w_2}$ (Figure~\ref{fig:logical_plans}b). During pair-wise join replication (step~2), Nova uses the join matrix $M$ to create one replica for each join pair, yielding two replicas $j_1, j_2$ for ${t_1,t_2} \Join w_1$ in Region~1 and $j_3, j_4$ for ${t_3,t_4} \Join w_2$ in Region~2, as shown in the parallelized plan $\Omega'_{log}$ (Figure~\ref{fig:logical_plans}b). Finally, in the virtual placement step, Nova computes for each replica the geometric median of the embedded coordinates of its connected nodes (i.e., the two sources and the sink). For example, the replica joining $t_1$ and $w_1$ minimizes the sum of distances to $\{\hat{t}_1, \hat{w}_1, \hat{\text{sink}}\}$ in the cost space. The resulting geometric median locations, shown as triangles~$\blacktriangle$ in Figure~\ref{fig:nova_overview}a, represent the replicas’ ideal positions before constraints are applied.

\subsection{Physical Replica Assignment}
\label{sec:nova_placement}
In the final phase, Nova iterates over the operators in $\Omega'_{log}$. Pinned operators (e.g., source, sink) are placed on their predefined nodes (Algorithm~\ref{alg:nova}, line 7), while join operators (line 10) are parallelized.
This phase addresses physical placement by mapping virtual operators to physical nodes under resource limits. Unlike NEMO~\cite{chatziliadis2024nemo}, which reduces rates through decomposable aggregations, joins amplify traffic due to replication and key-based repartitioning and thus require different algorithms. Nova introduces per-link bandwidth constraints alongside node-capacity limits, decomposes joins into key-local and region-aligned sub-joins, and applies bandwidth-aware partitioning to size replicas. It selects candidate nodes with a $k$-NN search in cost space and places replicas only if node capacity and link bandwidth permit. This approach yields horizontal replication near sources, which is fundamentally different from NEMO’s vertical aggregation trees. The following subsections explain stream partitioning, candidate selection, and replica placement.

\textbf{Bandwidth-Aware Stream Partitioning.}
Nova decomposes the left ($l_l$) and right ($r_l$) input streams of a join operator $\omega_j$  into disjoint partitions, creating multiple instances of $\omega_j \mapsto\{ \omega_{j_1}', \dots, \omega_{j_\rho}' \}$ with reduced resource requirements to satisfy the capacity constraint (Equation~\ref{eq:const_capcon}).
For instance, consider a join operator $\omega_j$ with $dr(l_l) = 25$ and $dr(r_l) = 25$ results in $C_r(\omega_j) = 50$. By partitioning each stream into 25 sub-streams, i.e., $l_l \mapsto \{ l_{l_1}', \dots, l_{l_{25}}' \}$ and $r_l \mapsto \{ r_{l_1}', \dots, r_{l_{25}}' \}$, the join can be parallelized across $25 \times 25 = 625$ replicas, where each replica $\omega_{j_i}'$ has $C_r(\omega_{j_i}')=2$. However, excessive partitioning increases network traffic from $50$ tuples/sec (no partitioning) to $625 \times 2 = 1250$ tuples/sec (maximum partitioning), a $25\times$ increase that can violate the bandwidth constraints (Equation~\ref{eq:const_bw}).

To adhere to bandwidth constraints, Nova introduces a scaling factor $\sigma$ $(0 \leq \sigma \leq 1)$ that controls the degree of partitioning.
Smaller $\sigma$ increases partitioning, improving robustness to overload but increasing bandwidth overhead and Phase III complexity, while larger $\sigma$ reduces these costs at the expense of higher overload risk.

Using $\sigma$, we define a maximum partition load threshold $p_{max}$ for streams $s$ and $t$ as:

\begin{equation}
    p_{max}(s, t) = \max(1, \sigma \cdot 0.5 (dr(s) + dr(t)))
\end{equation}

Each stream is then decomposed into partitions such that no partition exceeds $p_{\max}$ in data rate, ensuring that the resulting join replicas satisfy the capacity constraint (Equation~\ref{eq:const_capcon}). Assigning an equal weight of $0.5$ to each data rate relative to $\sigma$ improves resource utilization and reduces data movement compared to partitioning streams independently based on $\sigma$.

For example, consider a join operator $\omega_j$ with input streams $s$ and $t$, where $dr(s) = 2$, $dr(t) = 10$, and $\sigma = 0.5$. Independent partitioning yields $s \mapsto \{s_1', s_2'\}$, with $dr(s_1') =dr(s_2') = 1$, and $t \mapsto \{t_1', t_2'\}$ with $dr(t_1') =dr(t_2') = 5$, requiring $C_r(\omega_{j_i}') = 6$ for each join replica $\omega_{j_i}'$ of $\omega_j$, and a total transfer of 24 tuples/sec.
In contrast, our weighting sets $p_{max}(s, t) = \max(1, 0.5 \cdot 0.5 \cdot 12) = 3$, leaving $s$ unpartitioned ($s \mapsto \{s'\}$) while splitting $t$ into $t \mapsto \{t_1', \dots, t_4'\}$, with $dr(t_i') = 3$ for $t_i' \in \{t_1', t_2', t_3' \}$, and $dr(t_4') = 1$.
This strategy reduces the individual resource requirements to $C_r(\omega_{j_i}') = 5$ for the replicas $\omega_{j_i}'$ joining $s \Join \{t_1', t_2', t_3'\}$) and $C_r(\omega_{j_i}') = 3$ for $s \Join t_4'$, while reducing the network transfer rate to 18 tuples/sec.

The scaling factor $\sigma$ can be derived based on the data rates between streams $s$ and $t$, subject to a bandwidth constraint $t_b$, by solving the following convex optimization problem:

\begin{equation}
    \underset{\sigma \in [0, 1]}{\mathrm{arg~min}} (\sigma \cdot 2 \cdot dr(s) \cdot dr(t) - t_b)^2.
\end{equation}

Alternatively, operators may manually tune $\sigma$ rather than strictly adhering to bandwidth constraints. As network load grows exponentially with increasing partition count, balancing partitioning efficiency with bandwidth utilization often yields better performance. 
In our experiments, we set $\sigma = 0.4$, which our empirical analysis showed provides a well-balanced trade-off across diverse workloads and topologies.

\textbf{Candidate Selection.} After partitioning a join operator’s input streams, Nova places replicas on nodes to execute the decomposed joins. For each operator $\omega \in \Omega'_{\text{log}}$, Nova selects candidate nodes $V_{\text{knn}}(\omega) \subseteq V$ via a $k$-nearest-neighbors (kNN) search in the cost space, using an exact neighborhood search index (e.g., a k-d tree) for small topologies and an approximate Annoy-based index~\cite{bernhardsson1annoy} for large ones. The $k$ nodes whose coordinates are closest to the operator’s virtual coordinates $\hat{\omega}$ are selected as candidates.
To determine \(k\), Nova computes the ratio of an operator’s total required capacity to the median of available capacity per node. This ensures that the number of candidates considered for each operator automatically scales with its workload demand. 
Importantly, Nova does not attempt to use all available nodes. Only nodes that satisfy the compute and bandwidth constraints are considered as candidates, and replicas are placed on as many nodes as needed to avoid overload. Unused nodes remain idle.

\textbf{Replication and Placement.} In the final step, Nova assigns the partitioned input
$l_l \mapsto \{l_{l_1}', \dots, l_{l_m}'\}$ and $r_l \mapsto \{r_{r_1}', \dots, r_{r_n}'\}$ of a join operator $\omega_j$ to $m \times n$ replicas,
$\omega_j \mapsto \{\omega_{j_{11}}', \dots, \omega_{j_{mn}}'\}$.
Each replica $\omega_{j_i}'$ of $\omega_j$ is then placed sequentially on a node
$\nu_i \in V_{\text{knn}}(\omega_j)$ that satisfies $C_r(\omega_{j_i}') \leq C_a(\nu_i)$ 
(Equation~\ref{eq:const_capcon}).  
If no node meets this requirement, Nova either (1) distributes the remaining replicas evenly across $V_{\text{knn}}(\omega_j)$, accepting a risk of overload, or (2) expands $V_{\text{knn}}(\omega_j)$ to include additional nodes, potentially increasing network overhead.

\textbf{Example.} Figure~\ref{fig:nova_overview}b shows the parallelization and placement of the join operators $\omega_j \mapsto j_1, \dots, j_4$ in the cost space for the provided example (cf. Section~\ref{sec:nova_example}), where all source streams have a data rate of $25$. We use a $k$-NN search with $k=2$, $C_{\text{min}}=15$, and $\sigma=0$. In the following we explain how Nova places these replicas. For clarity, $\nu|C_a(\nu)$ denotes a node $\nu$ with available capacity $C_a(\nu)$, and $\omega|C_r(\omega)$ denotes an operator $\omega$ with required capacity $C_r(\omega)$. 
The search for $j_1$ returns $V_{\text{knn}}({j_1}) = [A|55, B|40]$. Since $A|55 \ge {j_1}|50$, operator ${j_1}$ is placed on node $A$. 
The search for ${j_2}$ returns $V_{\text{knn}}({j_2}) = [B|40, C|40]$, as $C_a(A)$ is now below $C_{\text{min}}$. Since $B|40 \le {j_2}|50$, the operator ${j_2}$ is replicated 625 times due to $p_{\text{max}}=1$, each representing a join between the partitions $t_{2_1} \Join w_{1_1}, \dots,t_{2_{25}} \Join w_{1_{25}}$. Half of the join replicas are placed and merged on node $B$ and the other half on node $C$.
For ${j_3}$ and ${j_4}$, the searches return $V_{\text{knn}}({j_3}) = [G|200, F|20]$ and $V_{\text{knn}}({j_4}) = [G|150, F|20]$, resulting in both operators being placed on node $G$.

Figure~\ref{fig:logical_plans}c shows the final logical plan $\Omega_{log}'$ generated by Nova. \\
Figure~\ref{fig:nova_overview}c depicts its placement in the cost space, illustrating the complete data flow from all sources $\{t_1, \dots, t_4\} \Join \{w_1, w_2\}$ to the sink, with the join distributed across nodes $A, B, C,$ and $G$.

\subsection{Re-optimization and Adaptivity}
\label{sec:nova_reopt}
Nova supports re-optimization without recomputing the full operator placement by distinguishing three types of changes: (1) adding sources or workers, i.e., nodes not yet involved in processing; (2) removing sources, join nodes, or workers; and (3) workload imbalances caused by changes in input rates or available resources. In NCS-based embeddings, changes in IP routes may alter latency estimates and shift node coordinates. To handle substantial changes, such as those introduced by mobile nodes~\cite{PazZM21}, Nova removes the affected nodes and re-adds them to ensure consistent embeddings. Nova focuses exclusively on optimizing operator placement and parallelization, independent of deployment, fault-tolerance and monitoring mechanisms, which are addressed by complementary work in volatile, geo-distributed environments~\cite{chaudhary2025ISQP, kozar2024fault, chatziliadis2021monitoring, burrell2023workload, chaudhary2026incremental}.

\textbf{Topology Changes.}
When a node is added, Nova first determines its role and then applies the corresponding re-optimization procedure. For a new worker node, Nova collects latency measurements from a fixed set of neighboring nodes in the NCS and computes the node’s coordinates in the cost space by minimizing the relative distance error, as described in Section~\ref{sec:nova_embedding}. Since the neighborhood size is constant, this computation has fixed time complexity. Nova then updates the neighborhood search index to make the node available for subsequent candidate selection. If the added node is a source, Nova updates the logical plan and the join matrix~$M$ (cf. Figure~\ref{fig:logical_plans}b) and re-executes Algorithm~\ref{alg:nova} only for the affected sub-branch.

Before removing a node, Nova again identifies its role. Idle workers are removed directly from the cost space and the neighborhood search index. If the node is a source, Nova deletes its entry from the join matrix~$M$ and removes all associated join replicas and dependencies from both the logical plan and the placement mapping. If the node hosts a join operator, Nova retrieves the operator’s precomputed virtual position in the cost space (cf. Section~\ref{sec:nova_vp}) and re-executes Phase~III to re-place and parallelize only the affected operator (cf. Section~\ref{sec:nova_placement}).

\textbf{Workload Changes.}
When workload characteristics change, Nova identifies the cause of the change and rebalances only the affected sub-joins. If source data rates change, Nova undeploys all upstream replicas associated with the affected sources and re-executes physical placement. If available resources on a worker change, Nova undeploys all operators running on that worker and re-executes physical placement for the affected replicas. Because the virtual placements in the cost space remain optimal under such changes, Nova skips virtual placement and recomputes only the physical placement (Phase~III) for the affected operators, as described in Section~\ref{sec:nova_placement}.

\begin{figure*}[t!]
  \centering
  \subfloat[FIT IoT Lab]{\includegraphics[scale=0.295]{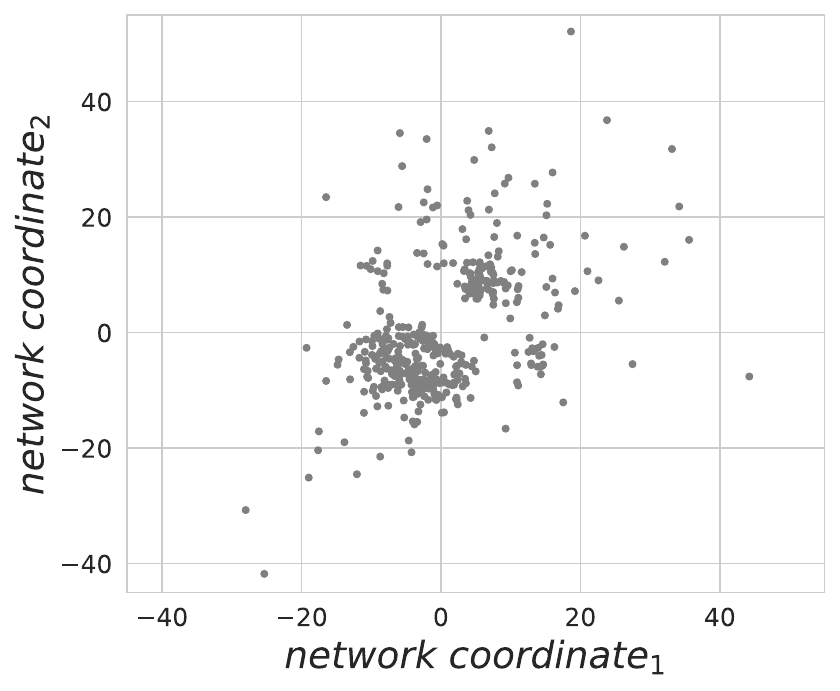}}
  \subfloat[PlanetLab]{\includegraphics[scale=0.295]{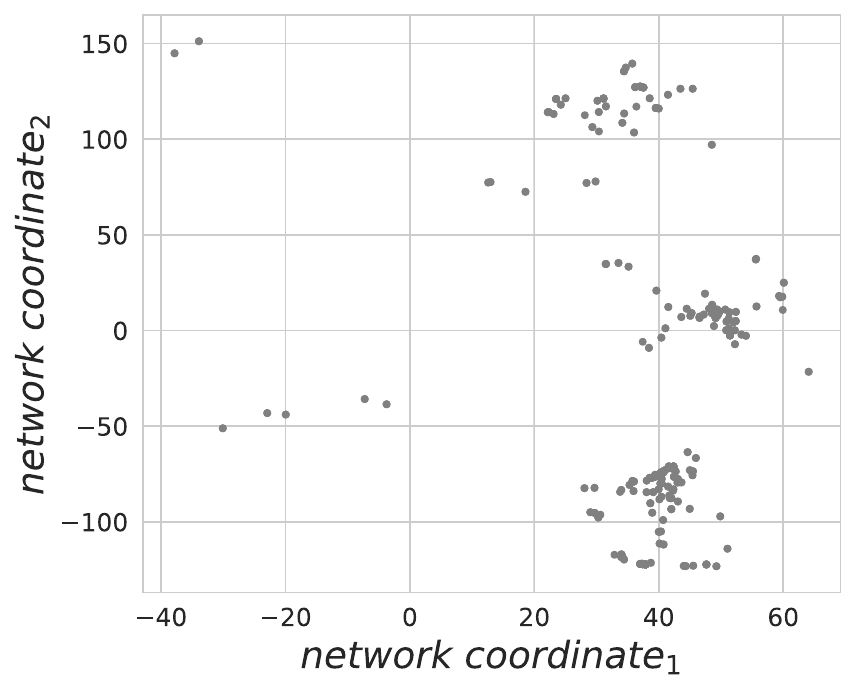}}
  \subfloat[RIPE Atlas]{\includegraphics[scale=0.295]{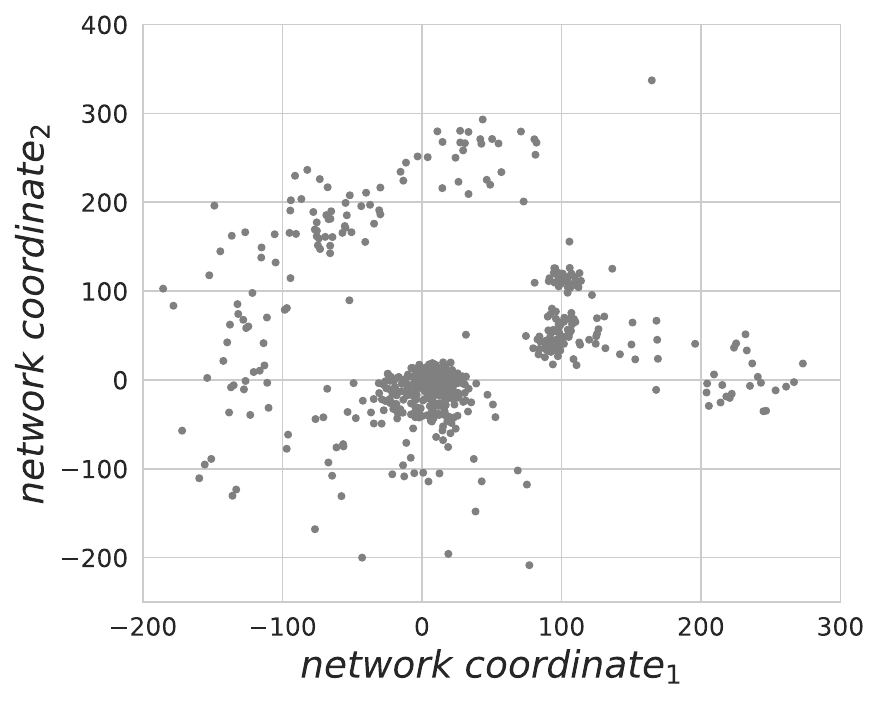}}
  \subfloat[King]{\includegraphics[scale=0.295]{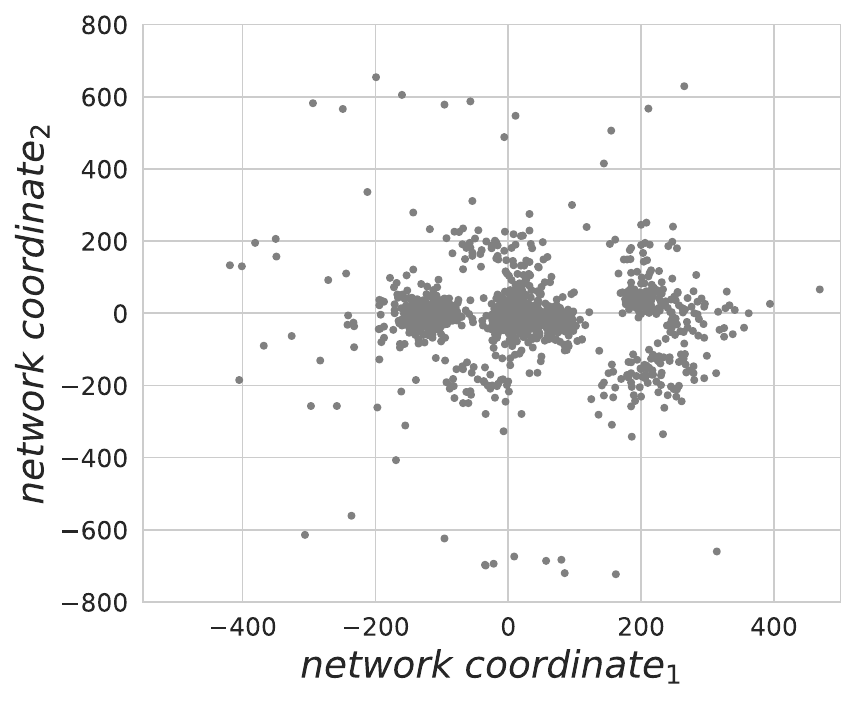}}
  \caption{Comparison of the network coordinate systems of the different topologies we used to evaluate Nova.}
  \label{fig:ncs_overview}
\end{figure*}

\subsection{Extensibility}
\label{sec:nova_extensions}

In this section, we describe modifications to Nova’s optimization to support additional cost metrics and extensions to support more complex workloads.

\textbf{Integrating Additional Cost Metrics.}
A key strength of Nova is that its cost-space formulation can be extended to support optimization over additional metrics beyond latency. Formally, Nova represents network characteristics through the symmetric matrix $A$ (cf.~Section~\ref{sec:nova_embedding}), which encodes pairwise distances between nodes and is embedded into a Euclidean cost space. Additional distance-based metrics, such as energy consumption or monetary cost, can be incorporated by augmenting $A$ with corresponding metric-specific distance matrices, following prior work by~\citet{PietzuchNAOP}. These additional matrices can be embedded as additional dimensions using techniques such as MDS~\cite{hout2013multidimensional} or Vivaldi~\cite{vivaldi04}, as described in Section~\ref{sec:nova_embedding}. 
During virtual placement (Phase~II, Section~\ref{sec:nova_vp}), operators are placed by minimizing distances in this augmented cost space, such that the optimization objective in Equation~\ref{eq:opt_func} implicitly balances latency with the added metrics without changing the structure of the optimization problem. 

Metrics that are not distance-based, such as memory capacity or specialized hardware availability, are integrated as constraints during physical placement (Phase~III, Section~\ref{sec:nova_placement}). Specifically, the candidate set $V_{\text{knn}}(\omega)$ is filtered to include only nodes that satisfy the corresponding resource constraints (cf.~Equations~\ref{eq:const_capcon}–\ref{eq:const_capmin}).

\textbf{Complex Workloads.}
Nova focuses primarily on the placement of partitionable, geo-distributed joins, but its concepts naturally extends to richer operator graphs that include filters, aggregations, and multi-way joins. 
Regarding joins over highly diverse key spaces, Nova mitigates skew through the join matrix, which specifies the sensors participating in each join. When joinable sensors exhibit high key-space diversity or high data rates, Nova further balances the load by partitioning the join as defined in Phase III. Joins over non-partitionable key spaces are outside Nova’s scope, as Nova fundamentally relies on partitioning and replication for scalability.

The key idea for complex workloads is to generalize the placement objective in Phase~II (Section~\ref{sec:nova_vp}) using a spring-force model, as proposed by~\citet{rizou2010op}, where springs represent communication costs between operators, yielding a convex optimization problem that remains efficiently solvable.
Simple, stateless filters can be colocated with upstream operators due to their negligible overhead. For aggregations, operators can be added to the spring-force model in Phase~II to determine their virtual placement within the same cost-space abstraction as NEMO~\cite{chatziliadis2024nemo}. Phase~III then parallelizes and places decomposable aggregations using NEMO.

For multi-way joins, the query can be decomposed into a sequence of two-way joins, each modeled as a node in the spring-force graph. Join-order optimization is orthogonal to Nova and can be addressed using existing work~\cite{ziehn2025unraveling}. However, since multi-way joins can produce exponentially growing outputs, Nova's bandwidth-aware partitioning helps manage intermediate data volumes by splitting streams proportionally to input rates and node capacities.

\section{Evaluation}
\label{sec:evaluation}

We evaluate Nova in two parts. First, we extend the simulations of~\citet{chatziliadis2024nemo}, using cost spaces derived from both real and synthetic topologies (Sections~\ref{sec:eval_load}--\ref{sec:eval_scalability}).
Second, we integrate Nova into NebulaStream’s optimizer~\cite{Zeuch2020nes, vliot_NES} and evaluate an end-to-end deployment on a physical cluster (Section~\ref{sec:eval_nova_deployment}).

\subsection{Experimental Setup}
\label{sec:eval_setup}

\quad \textbf{Hardware.} Simulations are executed using sequential Python scripts on a workstation with an AMD Ryzen 7 5700X CPU and 32 GB RAM. Following the experimental setup of NEMO~\cite{chatziliadis2024nemo}, we conduct our end-to-end experiments on the same testbed, consisting of a cluster of 14 Raspberry Pi 4B devices (Quad-Core Cortex-A72, 1.5 GHz, 4 GB RAM). All nodes run Ubuntu 22.04 and are interconnected via Gigabit Ethernet through a switch.

\textbf{Simulation.} We use the Vivaldi algorithm~\cite{vivaldi04} to generate network coordinate systems (NCSs) from latency measurements of real-world and artificial topologies.  
Specifically, we use latency measurements from the following topologies (cf. Figure~\ref{fig:ncs_overview}):  
1) \emph{FIT IoT Lab}~\cite{adjih2015fit}, an IoT testbed in France, where we evaluate round-trip times (RTTs) from 433 geographically distributed nodes, including microcontrollers, ARM devices, and four gateway servers.  
2) \emph{RIPE Atlas}~\cite{staff2015ripeAtlas}, an Internet measurement platform, where we analyze RTTs from 723 globally distributed anchors that serve as fixed latency measurement points.  
3) \emph{PlanetLab}~\cite{chun2003planetlab}, consisting of RTTs from 335 nodes hosted by universities and research institutions in Europe and North America.  
4) \emph{King}~\cite{gummadi2002king}, containing latency measurements from 1,740 Internet DNS servers.  

To select the number of neighbors $m$ for Vivaldi, we evaluate network coordinate accuracy using NCSIM~\cite{chen2011phoenix} and measure the mean absolute error (MAE). We observe that MAE converges quickly as \(m\) increases, with negligible accuracy improvements beyond a small neighborhood size. Our experiments thus lead to the same neighborhood size as prior work, i.e., \(m=20\) for RIPE Atlas and FIT IoT Lab, and \(m=32\) for PlanetLab and King.

To evaluate Nova's scalability on large topologies beyond the scale of existing systems, we generate synthetic NCSs with varying latency distributions and sizes from $10^3$ to $10^6$ nodes. Nodes are positioned within $[0,100]$ (x-axis) and $[-50,50]$ (y-axis), using Gaussian clusters to emulate heterogeneous, geo-distributed networks. These synthetic topologies enable controlled scalability testing beyond real-world measurements (cf. Section~\ref{sec:eval_scalability}).

\textbf{End-to-end Deployment.} To evaluate Nova’s end-to-end performance, we integrate it into NebulaStream~\cite{Zeuch2020nes} for operator placement and deploy NebulaStream on the testbed. In this setup, one Raspberry Pi acts as the coordinator/sink, eight serve as sources, and the remaining nodes function as workers. To emulate real-world conditions of network and compute heterogeneity, we follow the methodology of NEMO~\cite{chatziliadis2024nemo}: injecting artificial latency between nodes based on RIPE Atlas data using Linux’s \texttt{tc} tool, and applying artificial load to the source nodes with \texttt{stress} (full CPU utilization and 80 \% memory usage).

\begin{figure}[t!]
  \includegraphics[width=\columnwidth]{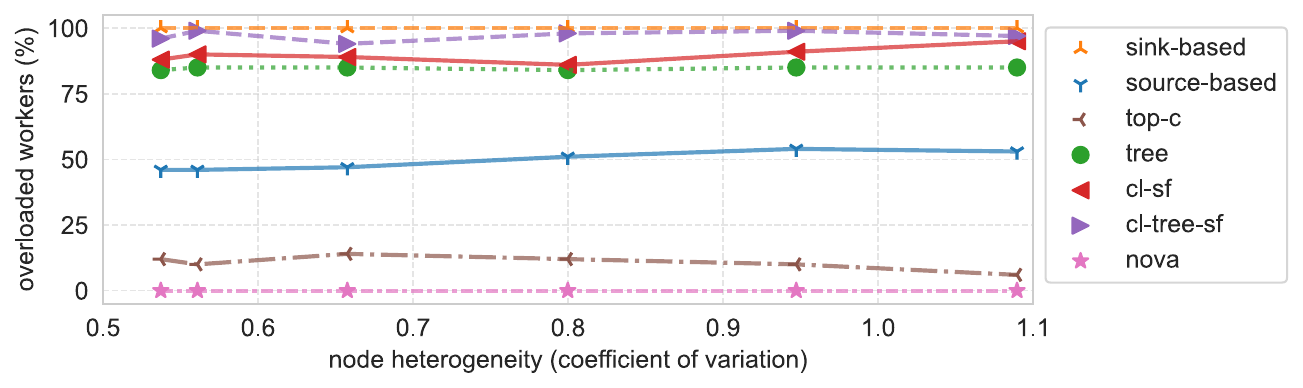}
  \caption{Percentage of overloaded nodes versus node heterogeneity (Coefficient of Variation of node capacities). Higher CV indicates greater resource imbalance.}
  \label{fig:nova_eval_load}
\end{figure}

\textbf{Workloads.} In our simulations, we evaluate Nova using geo-distributed join queries involving two conceptual streams, each originating from multiple joinable data sources. To assess performance under varying loads, we adjust the computational capacities, number of sources, and data rates. 
We control the heterogeneity of the topology by varying node capacities while keeping the total capacity approximately constant. 
Specifically, we transition from a near-uniform capacity distribution (low heterogeneity) to increasingly skewed distributions (high heterogeneity), ranging from uniform (capacities between 1 and 200, median around 35) to exponential (capacities between 1 and 1000, median around 28). 
These variations produce increasing coefficients of variation (CV), which we use to quantify node heterogeneity. 
Minor fluctuations in CV values may occur due to topology size and distribution sampling.
In addition to node-capacity heterogeneity, we vary the per-source input rates to create load imbalances similar to those caused by skewed key distributions.
We randomly designate 60\% of the nodes as sources and the remaining 40\% as workers, mirroring the hardware distribution in the FIT IoT Lab~\cite{adjih2015fit}. As the topology grows, the increasing number of sources proportionally raises the overall load. 
The sink is randomly selected to avoid bias. 
Since Nova is designed for 2-way joins, each source is randomly assigned to one of two logical streams and joined with exactly one other source, such that the corresponding join matrix contains exactly one entry per row. Finally, we uniformly distribute source data rates between 1 and 200 to represent varying resource requirements. 

We evaluate Nova in our end-to-end deployments using an environmental monitoring workload inspired by the DEBS~2021 Grand Challenge~\cite{tahir2021debs}. We join pressure and humidity readings from four regions using tumbling windows ranging from 1 ms to 1 s, resulting in four parallel two-way joins across eight sensors. Each sensor generates data at a fixed rate of 1 kHz (1000 tuples per second), such that source windows contain between one tuple for 1 ms windows and 1000 tuples for 1 s windows. 

We intentionally include very small windows (e.g., 1 ms) to induce high operational pressure in our small-scale testbed, rather than to reflect typical application workloads. Such window sizes substantially increase the frequency of window creation, watermark propagation, and state and buffer management, leading to significant system overhead. This synthetic load enables us to approximate the behavior of much larger deployments, where comparable overhead arises from many concurrent sources, without requiring a proportional increase in physical nodes.
At the other extreme, larger windows (e.g., 1 s) shift the bottleneck toward per-window processing and buffering costs.

\textbf{Baselines.} We compare Nova against the following baselines: 
1) Sink-based: The default approach of NebulaStream~\cite{Zeuch2020nes}, where joins are computed at the sink node.
2) Source-based: A locality-aware heuristic adapted for streaming joins, resolving the join matrix by placing each join at the source with the highest data rate~\cite{sundarmurthy2021locality}.
3) Top-c: A resource-aware heuristic representing cloud-based approaches by placing and parallelizing the join on the node with the highest computational capacity.
4) Tree: A WSN-based approach where the topology forms a minimum spanning tree (MST), joining data at intersections~\cite{mihaylov2010dynamicjoinvldb}.
5) Cl-SF: An adapted WSN approach that clusters the topology using LEACH-SF~\cite{shokouhifar2017leachsf}, computing joins at intersecting cluster heads or the sink if none exist.
6) Cl-Tree-SF: A hybrid of Tree and Cl-SF, clustering the topology, forming an MST among cluster heads, and computing joins at intersections.

\begin{figure}[t!]
    \centering
    \includegraphics[width=\columnwidth]{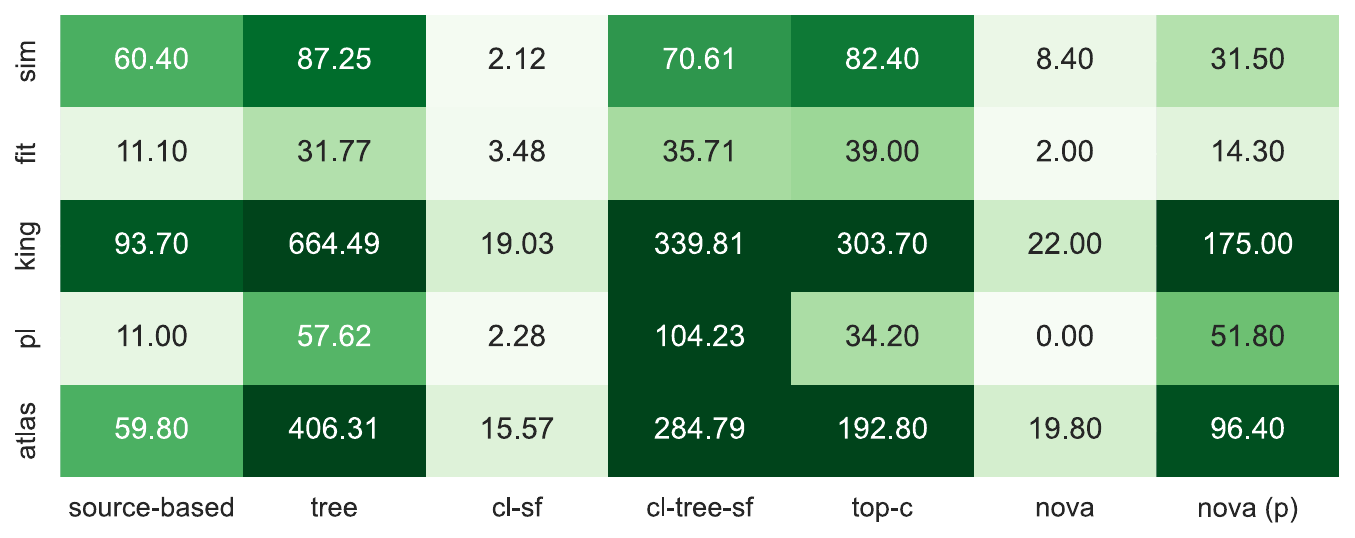}
    \caption{Latency deltas (90th percentile) relative to sink-based direct transmission across different approaches.}
    \label{fig:eval_lat_heatmap}
\end{figure}

\subsection{Mitigation of Over-Utilization}
\label{sec:eval_load}
In this section, we evaluate Nova's ability to avoid node over-utilization under increasing heterogeneity. Figure~\ref{fig:nova_eval_load} shows the percentage of overloaded nodes compared to six baselines, under varying node heterogeneity, quantified by the CV of node capacities (cf. Section~\ref{sec:eval_setup}). Higher CV values indicate greater heterogeneity and resource-imbalance. 
We report overloaded nodes, defined as those exceeding their capacity, as a percentage of total workers.
We focus on results from a 1000-node synthetic topology, as trends remained consistent across different scales and topologies.

As shown in Figure~\ref{fig:nova_eval_load}, Nova consistently outperforms all baselines, maintaining zero overloaded nodes across all capacity distributions. In contrast, the sink-based approach performs the worst, invariably overloading 100\% of its workers since the sink node alone handles all computation.
Among the remaining baselines, the cluster- and tree-based approaches used by WSNs show the highest overload, with the hybrid variant (Cl-Tree-SF) ranging from 94\% to 99\% overloaded nodes, followed by Cl-SF at 86\% to 95\%, and the pure tree-based approach at roughly 85\% across all distributions. 
Because these methods are resource-agnostic, they concentrate the join workload on a limited set of nodes chosen primarily for their topological position, rather than their capacity.

The source-based approach shows similar behavior but nearly halves the overload compared to WSN-based methods by distributing computation among more source nodes. Nonetheless, it remains resource-agnostic, resulting in 46\% overload in the homogeneous scenario and rising to 54\% under higher heterogeneity.
The best-performing baseline is top-c (6\%-14\% overloaded nodes), as it assigns sub-joins to the node with the highest available capacity, but lacks distributed parallelization, so large computations can still overwhelm individual nodes.

In summary, Nova’s resource awareness, combined with parallelization and partitioning, prevents worker over-utilization across capacity distributions, significantly outperforming all baselines.

\subsection{Placement Quality}
\label{sec:eval_latency}
In this section, we evaluate the theoretical latencies of each approach by comparing their 90th percentile (90P) latencies (in milliseconds) against the theoretical lower bound defined by the sink-based solution (Figure~\ref{fig:eval_lat_heatmap}). For clarity, we exclude processing delays and estimation errors in this analysis. The effects of estimation errors are discussed in Section~\ref{sec:eval_estimation_error}, while end-to-end latencies in real-world deployments are evaluated in Section~\ref{sec:eval_nova_deployment}.

Overall, Nova and Cl-SF achieve the lowest latencies, consistently approaching the theoretical lower bound across all topologies. Cl-SF shows minimal deviations from the lower bound, with deltas of 2–3.5 ms for PlanetLab, FIT, and 1K-node simulations, 15.57 ms for RIPE Atlas, and 19.03 ms for King, thanks to its clustering design that prioritizes minimal latency. Similarly, Nova achieves near-optimal latencies (deltas of 0 ms on PlanetLab, 2 ms on FIT, 8.4 ms on the 1K simulation, 19.8 ms on RIPE Atlas, 22 ms on King) when workloads can be accommodated on a small number of nodes. 
Nova(p) denotes the placement generated for the most heterogeneous topology, which requires the highest degree of replication to achieve load balance.
In this challenging setting, Nova(p)'s latencies increase up to 14.3 ms (FIT), 31.5 ms (1K simulation), 51.8 ms (PlanetLab), 96.4 ms (RIPE Atlas), and 175 ms (King).

Compared to Nova and Cl-SF, source-based and top-c approaches exhibit moderately higher latencies, often exceeding 10 ms for PlanetLab and FIT, and reaching up to 190+ ms for RIPE Atlas and King. However, they still outperform tree-based methods (Tree, Cl-Tree-SF), which show significantly larger latencies due to multi-hop routing. For instance, tree-based methods exhibit deltas ranging from 35 ms (FIT) to over 650 ms (King), which are more than twice as high as Nova's fully parallelized placements (Nova(p)).

In summary, Nova’s resource-aware and parallelized design allows it to operate close to the theoretical lower bound when feasible. It also scales gracefully under heavier workloads by leveraging operator replication and stream partitioning to prevent overload, while maintaining low latency.

\begin{figure}[t!]
    \centering
    \includegraphics[width=\linewidth]{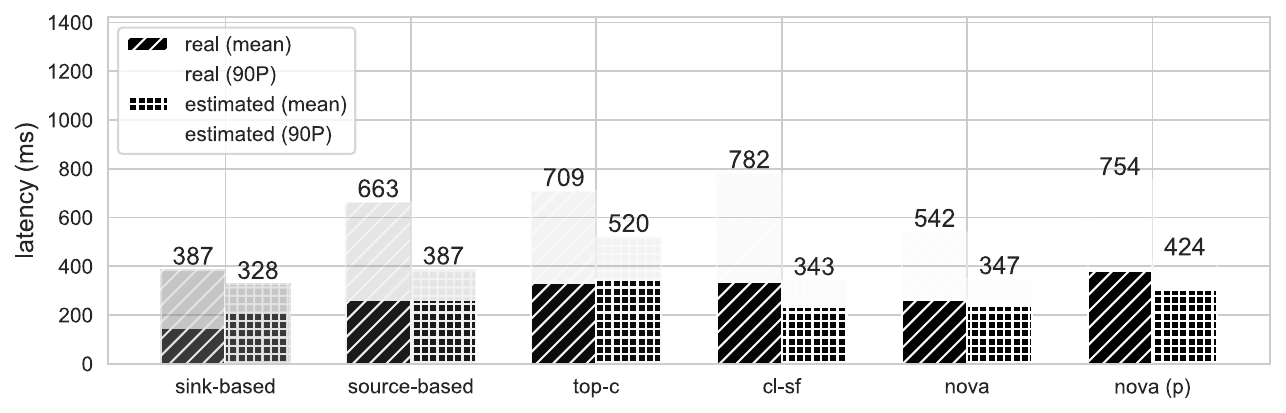}
    \caption{Mean and 90th percentile latencies using NCS estimates (left) vs. real measurements (right) on RIPE Atlas.}
    \label{fig:lat_comparisons}
\end{figure}

\subsection{Impact of Estimation Errors}
\label{sec:eval_estimation_error}

This section evaluates the impact of NCS estimation errors by comparing estimated and real mean and 90th percentile latencies using measurements from a 418 node RIPE Atlas subset. Figure~\ref{fig:lat_comparisons} summarizes the results, with Tree and Cl-Tree-SF omitted because their latencies are orders of magnitude higher than all other approaches.

For mean latency, Nova exhibits low estimation error with 237 ms estimated versus 259 ms measured. Source-based and top-c also show small discrepancies at 261 ms versus 262 ms and 345 ms versus 330 ms, while sink-based shows a larger deviation at 220 ms versus 146 ms. In contrast, tree-based approaches suffer extreme underestimation, with Tree increasing from 512 ms to 11.7 s and Cl-Tree-SF from 450 ms to 2 s.

At the 90th percentile, tail effects increase estimation error across all approaches, with Nova rising from 347 ms to 542 ms and Nova(p) from 424 ms to 754 ms. Source-based increases from 387 ms to 663 ms and top-c from 520 ms to 709 ms, while tree-based approaches again exhibit extreme underestimation, jumping from 733 ms to 19 s. Overall, WSN-based approaches that are not cost-space optimized are highly sensitive to TIVs, whereas Nova maintains accurate mean estimates and competitive tail latency, achieving a real 90th percentile latency of 542 ms that is 35 times lower than tree-based approaches and comparable to top-c, demonstrating that cost-space optimization yields deployments robust to NCS estimation errors.

\begin{figure}[t!]
  \includegraphics[width=\columnwidth]{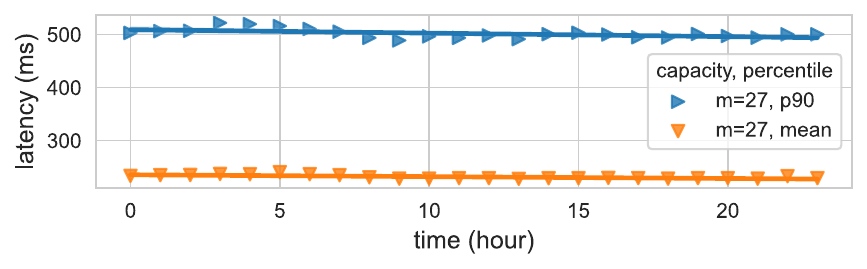}
  \caption{Mean and 90th percentile latencies of Nova for RIPE Atlas over 24 hours.}
  \label{fig:nova_eval_robustness}
\end{figure}

\subsection{Resilience to Latency Variations}
\label{sec:eval_evolution}
To evaluate Nova’s resilience to latency variations, we measured mean and 90P latencies over a 24-hour period on a fixed 418-node subset of RIPE Atlas.
We show only the results for the highest heterogeneity and level of parallelization with a median of 27 ms, as other distributions showed similar patterns. 
During this period, the number of changed latency entries between successive measurements over a 10 ms threshold ranged from 7k to 14k, with a median change magnitude of 24 ms.

Figure~\ref{fig:nova_eval_robustness} shows that the 90P latencies range from approximately 489 ms to 522 ms, while the mean latencies stay between 228 ms and 241 ms. Despite day-to-night transitions and varying network conditions, the observed standard deviations remain within tens of milliseconds for both metrics, indicating that Nova’s placements are largely unaffected by transient latency spikes or routing changes. These results highlight Nova’s robustness in long-running stream processing scenarios, reducing the need for frequent re-optimizations even under dynamic network conditions.

\subsection{Placement and Re-optimization Scalability}
\label{sec:eval_scalability}
\begin{figure}[t!]
  \includegraphics[width=\columnwidth]{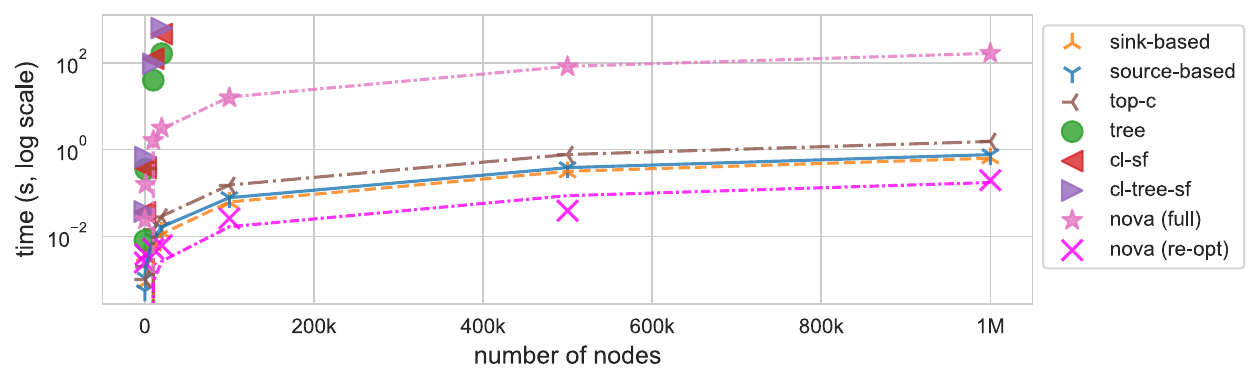}
  \caption{Optimization and re-optimization times (log scale) for workloads where both topology size and query complexity (number of operators and streams) grow proportionally.}
  \label{fig:nova_eval_scalability}
\end{figure}

In this section, we evaluate Nova’s scalability for full optimization and re-optimization. For each topology size, we trigger five re-optimization events per run: adding a source, removing a source, removing a worker, updating a node's coordinates, and changing the data rate of a source. Each event is applied to a single randomly selected node. To ensure comparability, the number of updates remains constant across all topology sizes, while query complexity grows proportionally with topology size (i.e., the number of operators and streams increases). Figure~\ref{fig:nova_eval_scalability} shows optimization time versus topology size, with the x-axis showing the number of nodes, and y-axis the cumulative runtime in seconds (log scale).

Nova scales efficiently, placing operators in topologies of up to 1 million nodes in approximately 135 seconds. Runtime grows consistently with topology size, increasing from 0.015 seconds at 100 nodes to around 135 seconds at 1 million nodes. While the underlying complexity remains near-linear, runtimes span several orders of magnitude, motivating the use of a log scale for clearer visualization. In contrast, Tree, Cl-SF, and Cl-Tree-SF baselines exceed the 10-minute timeout beyond 20,000 nodes, making them impractical for large-scale deployments.
While sink-based, source-based, and top-c methods remain consistently fast (under 2 seconds even for 1 million nodes), they ignore node capacity constraints, which leads to overloading. Nova, in contrast, ensures resource-aware placement at competitive runtimes.

Nova’s re-optimization times remain below one second even at 1 million nodes. Adding a new source, which is the most expensive event, takes 0.2 s, while all other re-optimization events complete in well under 0.1 s. This demonstrates Nova’s efficiency in adapting to dynamic changes without full recomputation.
In contrast, the baselines require full placement recomputation for the same events, as they lack incremental update mechanisms.

\subsection{End-to-end Performance}
\label{sec:eval_nova_deployment}
This section evaluates Nova's performance on the DEBS 2021 benchmark workload using a 14-node Raspberry Pi cluster as proposed by~\citet{chatziliadis2024nemo}. We measure end-to-end latency (network and computational delays) and throughput (total tuples processed) over a two-minute runtime, comparing Nova against our baselines. To ensure comparability, node churn is disabled.
In this topology, the cluster-based approaches cl-sf and cl-tree-sf, as well as top-c, produce identical placements and are grouped together. Similarly, the source-based and the tree approach yield the same placement and are also grouped together.

\begin{figure}[t!]
    \centering
    \includegraphics[width=\linewidth]{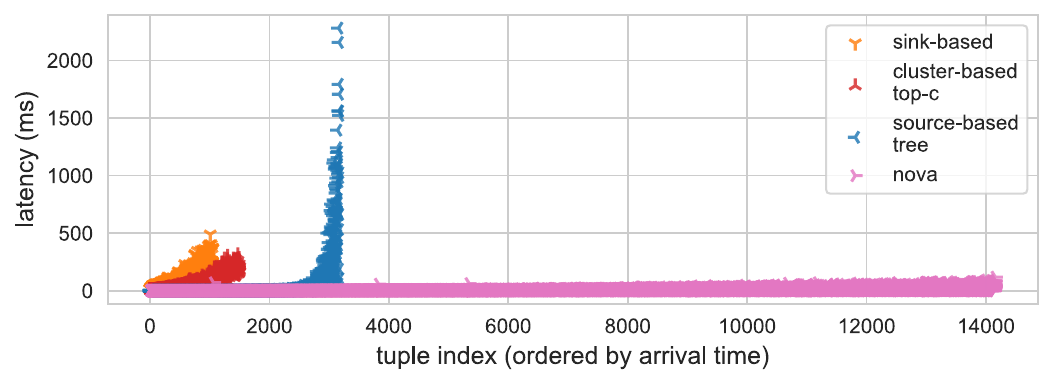}
    \caption{Latency trends over processed tuples for the DEBS 2021 end-to-end deployment (non-stressed)}
    \label{fig:e2e_latency_scatter}
\end{figure}

\textbf{Throughput.} Figure~\ref{fig:e2e_latency_scatter} illustrates Nova’s throughput advantage by depicting the number of processed tuples within the two-minute query runtime (x-axis) versus their latency (y-axis) for the non-stressed workload.
The sink-based approach achieves the lowest throughput, processing 1,057 tuples, as it performs all computations centrally, overloading the sink node. The cluster-based and top-c approaches achieve slightly higher throughput (1,503 tuples) by computing joins on a single cluster head, which has more resources than the sink but remains a bottleneck.

The source-based/tree baselines process 3,176 tuples, doubling throughput by distributing joins across multiple source nodes. However, these nodes lack computational capacity, as they share resources with data ingestion, leading to overloads.
In contrast, Nova achieves the highest throughput, processing 14,159 tuples, 4.5$\times$ more than the source-based/tree and 13.4$\times$ more than the sink-based approach. This order-of-magnitude improvement stems from Nova’s ability to parallelize join operations, distributing them across four worker nodes to prevent overload.

\begin{figure}[t!]
    \centering
    \includegraphics[width=\linewidth]{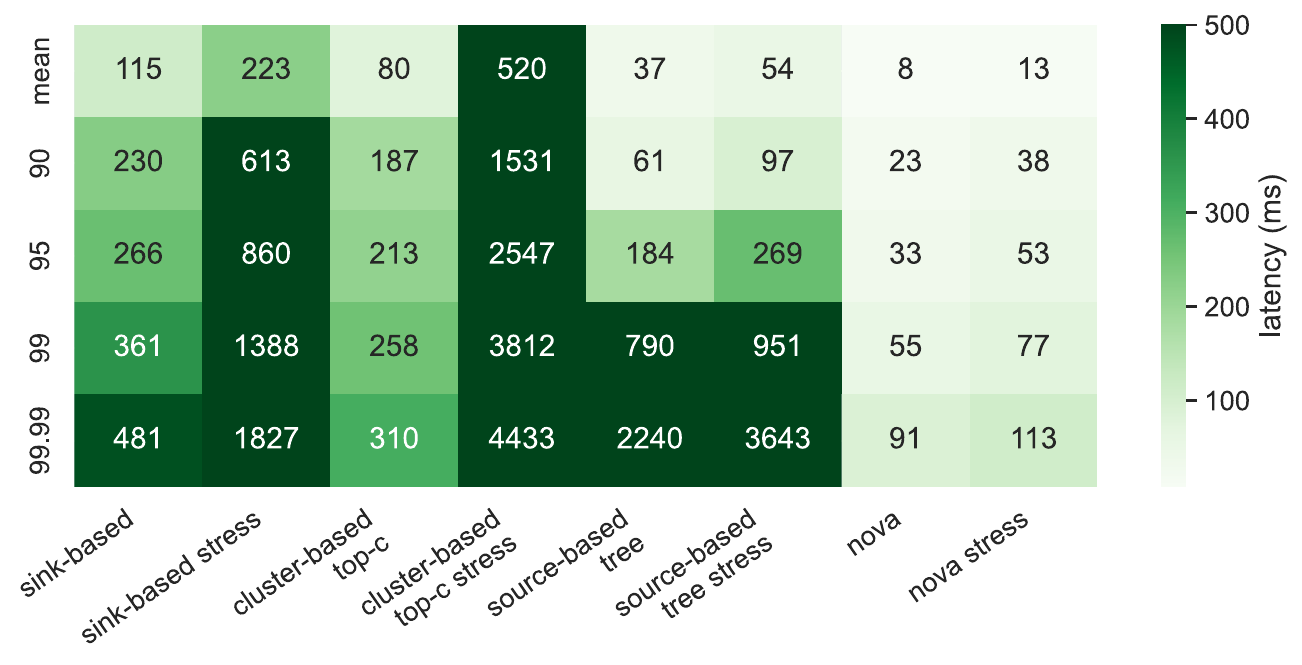}
    \caption{Comparison of the mean and 90-99.99 percentile latencies for the DEBS 2021 end-to-end deployment.}
    \label{fig:e2e_latency_percentiles}
\end{figure}

\textbf{Latency.} Nova’s parallelized execution and resource-aware placement are also reflected in the latency distributions, shown in Figure~\ref{fig:e2e_latency_percentiles}, which presents latency under both normal and stress conditions.
Nova achieves a mean latency of 8 ms, significantly outperforming the sink-based (115 ms, 14.4$\times$ higher), cluster-based/top-c (80 ms, 10$\times$ higher), and source-based/tree (37 ms, 4.6$\times$ higher) approaches. Nova’s 99.99th percentile latency remains tightly bounded at 91 ms, significantly lower than the sink-based (481 ms, 5.3$\times$ higher) and cluster-based/top-c (310 ms, 3.4$\times$ higher) baselines.

Under stress, Nova remains robust, with its mean latency increasing to 13 ms and a 99.99th percentile of 113 ms. In contrast, the cluster-based/top-c approach spikes to 4,433 ms at the 99.99th percentile (39$\times$ higher than Nova), while the source-based/tree approach reaches a mean latency of 54 ms (4.2$\times$ higher than Nova).

In summary, Nova achieves orders-of-magnitude improvements in both throughput and latency by optimizing the parallelization and placement of join operators to prevent node overload and minimize network delay. Its throughput surpasses the best baseline by 4.5$\times$, while its latency is 4.6–14.4$\times$ lower under normal conditions and 4.2–39$\times$ lower under stress. These results validate Nova’s suitability for resource-constrained, geo-distributed environments.

\section{Related Work}
\label{sec:related_work}

We group existing approaches into three main categories: WSN-based methods, cloud-based methods, and wide-area (WA)-based methods (including NEMO~\cite{chatziliadis2024nemo}). Although each category addresses crucial aspects of distributed stream processing, such as in-network data aggregation, elasticity, and operator scheduling, most fail to unify placement, partitioning, and replication in a way that is both resource-aware and capable of re-optimizing at large scales. In contrast, Nova addresses capacity constraints by parallelizing joins in an integrated fashion while being scalable and adaptable.

\textbf{WSN-based Methods.}
Early work in WSNs emphasizes reducing energy use and bandwidth through in-network aggregation. Cluster-based protocols (e.g., LEACH~\cite{leach}, HEED~\cite{heed}, CLUDDA~\cite{cludda}) select cluster heads that summarize data locally. Tree-based methods like EADAT~\cite{eadat} or PEDAP~\cite{pedap} construct an MST to minimize transmissions, while chain-based solutions such as PEGASIS~\cite{pegasis} organize nodes in a linear chain for incremental aggregation. These protocols effectively conserve energy but do not account for latency and node overloading.

Beyond aggregation, various WSN-oriented approaches tackle joins.
Multi-hop in-network joins~\cite{mihaylov2008substrate, mihaylov2010dynamicjoinvldb} optimize communication by decomposing join operations into partial computations distributed across network paths.
Broadcast~\cite{yu2006network, yang2007network} and hash-based~\cite{pandit2006communication, gupta2007communication} joins localize semijoin computations near data sources.
Filtering methods such as predicate embedding~\cite{abadi2005reed} and fragmented semijoins~\cite{kang2015network} reduce network traffic by discarding irrelevant readings locally and eliminating low-frequency data.
While these techniques parallelize workloads and reduce communication costs through partial in-network computation, they are primarily designed for small-scale topologies and do not effectively handle node capacity constraints, latency requirements, or overload in larger heterogeneous environments.

\textbf{Cloud-based Methods.}
Many cloud-based SPEs elastically adjust operator parallelism to handle workload fluctuations~\cite{Dhalion, xu2016stela, kalim2018henge, fu2017drs, kalavri2018three}. Streambed~\cite{rosinosky2024streambed} focuses on capacity planning for distributed dataflow, while~\citet{das2014adaptive} tune batch sizes to balance throughput and latency. Although effective in cloud settings, these approaches generally assume abundant resources or easy horizontal scaling and thus overlook tight capacity constraints, complex routing paths, and node heterogeneity.

Scheduling frameworks~\cite{ousterhout2013sparrow, xu2021move} and dynamic partitioning techniques~\cite{gedik2014partitioning,katsipoulakis2017holistic,nasir2016two} aim to meet latency targets and mitigate skew by distributing tasks and splitting streams. However, they typically treat placement separately from partitioning or replication, which can cause node overload in geo-distributed or resource-constrained environments.

Existing placement algorithms are often too costly or inflexible for online use~\cite{cardellini2016op,cardellini2018optimalrep} and may require manual tuning~\cite{rubio2015user}. For example, \citet{HeinzeJRP0JF15} propose a bin-packing heuristic that minimizes latency violations but does not explicitly model computational resource constraints such as memory and processing throughput. CAPSys~\cite{wang2024capsys} addresses resource contention but lacks support for complex geo-distributed topologies and does not jointly handle partitioning and replication like Nova.

\textbf{WA-based Methods.}
SBON~\cite{PietzuchNAOP} and its multi-operator placement extension~\cite{rizou2010op} apply cost-space embedding for iterative operator placement, yielding efficient solutions for single- and multi-operator scenarios, but prioritize placement optimization without incorporating constraints related to node capacities, replication strategies, or stream partitioning.
NEMO~\cite{chatziliadis2024nemo} addresses node capacities and replication for decomposable aggregation functions.
However, NEMO cannot process joins because (1) NEMO's tree-based aggregation decomposition reduces data volume at each tree level, but joins amplify data, making this decomposition inapplicable; (2) its optimization couples placement decisions across levels of the aggregation tree, whereas join replicas require independent per-replica optimization; and (3) its cost model does not account for bandwidth constraints and heterogeneous input rates, both critical for joins where operator replication increases network traffic.

Systems like WASP~\cite{jonathan2020wasp}, SWAN~\cite{song2022swan}, DART~\cite{liu2021dart}, Droplet~\cite{elgamal2018droplet} use ILP, dynamic programming, or heuristics to minimize end-to-end latency, job completion times, or bandwidth usage across data centers. Some solutions (e.g., WASP) focus on inter-data-center scheduling only, while others (e.g., DART) propose decentralized overlays for computation orchestration. While effective at reducing network overhead, most lack support for load reduction through operator replication or re-optimizations when the topology changes.

\section{Conclusion}
\label{sec:conclusion}
Nova provides a scalable and resource-aware approach for optimizing the placement and parallelization of join operators in geo-distributed environments. 
By framing placement as a convex geometric median problem in a Euclidean cost space and implementing bandwidth-aware stream partitioning, Nova efficiently approximates the NP-hard Operator Placement and Parallelization problem in linear time, while also enabling adaptivity in dynamic and heterogeneous environments.

Our results show that current latency-optimal strategies are insufficient, often overloading resource-constrained nodes when placing joins near data sources. Nova addresses this challenge by balancing load through joint optimization of placement, replication, and partitioning, thereby minimizing latency without overloading the network. 
Future work will empirically evaluate multi-way join performance, integrate predictive workload models, and explore different cost metrics beyond latency.

\begin{acks}
This work was funded by the German Federal Ministry for Education and Research as BIFOLD - Berlin Institute for the Foundations of
Learning and Data (ref. BIFOLD24B).
\end{acks}

\section*{Artifacts}
The code, datasets, and artifacts for Nova, together with instructions, are available at \url{https://github.com/xchatzil/NOVA}.

\balance
\bibliographystyle{ACM-Reference-Format}
\bibliography{references}

\end{document}